\documentclass[11pt,a4paper]{article}

\usepackage{microtype}
\usepackage{amsmath, amsfonts,amssymb,amsthm,mathtools,nicefrac,nccmath,cases,physics}
\usepackage{mathrsfs}  
\usepackage{bm}
\usepackage{soul}
\usepackage{dsfont}
\usepackage{booktabs,multirow}
\usepackage{tabularx}
\usepackage[usenames,dvipsnames,table]{xcolor}
\usepackage{colortbl}
\usepackage{caption}
\usepackage{subcaption}
\usepackage{float}
\usepackage{wrapfig}
\usepackage{hhline}
\usepackage{cite}
\numberwithin{equation}{section}
\numberwithin{table}{section}
\usepackage[linktocpage=true]{hyperref}
\hypersetup{colorlinks=true,linkcolor=colorloc3,citecolor=colorloc3,urlcolor=colorloc3}
\usepackage{afterpage}

\def\hybrid{\topmargin -20pt    \oddsidemargin 0pt
	\headheight 0pt \headsep 0pt
	\textwidth 6.5in        
	\textheight 9in         
	\textwidth 6.25in       
	\textheight 9 in       
	\marginparwidth .875in
	\parskip 5pt plus 1pt 
	\jot = 1.5ex
}
\hybrid

\definecolor{colorloc1}{RGB}{164,42,46} 
\definecolor{colorloc2}{RGB}{100,100,100} 
\definecolor{colorloc3}{RGB}{204,119,34}  
\definecolor{colorloc4}{RGB}{25,25,112}  
\definecolor{colorloc5}{RGB}{100,0,0}  
\definecolor{colorloc6}{RGB}{200,200,200}  
\definecolor{colorloc7}{RGB}{70,70,70}  
\definecolor{colorloc8}{RGB}{0,128,128}  

\usepackage[framemethod=default]{mdframed}

\newmdenv[skipabove=10pt,
skipbelow=7pt,
rightline=false,
leftline=true,
topline=false,
bottomline=false,
linecolor=colorloc4,
backgroundcolor=colorloc8!5,
innerleftmargin=4pt,
innerrightmargin=0pt,
innertopmargin=0pt,
leftmargin=2pt,
rightmargin=0pt,
linewidth=2pt,
innerbottommargin=4pt,
frametitlebackgroundcolor=colorloc3]{lbBox}

\newmdenv[skipabove=10pt,
skipbelow=7pt,
rightline=false,
leftline=true,
topline=false,
bottomline=false,
linecolor=colorloc5,
backgroundcolor=colorloc1!5,
innerleftmargin=4pt,
innerrightmargin=4pt,
innertopmargin=0pt,
leftmargin=2pt,
rightmargin=0pt,
linewidth=2pt,
innerbottommargin=4pt,
frametitlebackgroundcolor=colorloc1]{ldBox}

\usepackage{fancyhdr}

\usepackage{fancyhdr}
\usepackage{sectsty}
\usepackage{titlesec}

\usepackage[linktocpage=true]{hyperref}
\hypersetup{colorlinks=true,linkcolor=colorloc5,citecolor=colorloc4,urlcolor=colorloc4}

\usepackage[shortlabels]{enumitem}

\newcommand{\beq}{\begin{equation}}
\newcommand{\eeq}{\end{equation}}

\newcommand{\bea}{\begin{eqnarray}}
\newcommand{\eea}{\end{eqnarray}}
\newcommand{\be}{\begin{equation}}
\newcommand{\ee}{\end{equation}}

\begin{document}


	\baselineskip=14pt
	\parskip 3pt

	\vspace*{-1.5cm}
	
	\vspace{3cm}
	\begin{center}        

	{\bf \huge Stochastic inflation as a superfluid
}   
	\end{center}
	\renewcommand{\thefootnote}{\fnsymbol{footnote}}
	\vspace{0.2cm}
	\begin{center}        
		{ \large Gianmassimo Tasinato$^{a, b}$}\footnote{\texttt{g.tasinato2208@gmail.com}}
	\end{center}
	
	\begin{center}  
		\emph{\textsl{$^{a}$ Physics Department, Swansea University, SA28PP, United Kingdom}\\
\textsl{$^{b}$ Dipartimento di Fisica e Astronomia, Universit\`a di Bologna,\\
 INFN, Sezione di Bologna,  viale B. Pichat 6/2, 40127 Bologna,   Italy}}
	\end{center}
	
	\vspace{0.2cm}

	\begin{abstract}
		\noindent We point out that inflationary superhorizon fluctuations can be effectively described by a set of equations analogous to those governing a superfluid. This is achieved through a functional Schr\"odinger approach to the evolution of the inflationary wavefunction, combined with a suitable coarse-graining procedure to capture large-scale dynamics. The irrotational
fluid velocity is proportional to the gradient of the wavefunction phase. Marginalizing over short superhorizon modes introduces an external  force acting on the fluid velocity. The quantum pressure characteristic of the superfluid plays a  role in scenarios involving an ultra-slow-roll phase of inflation. Our superfluid framework is  consistent with the standard Starobinsky approach to stochastic inflation while offering complementary insights, particularly by providing more precise information on the phase of the inflationary wavefunction. We also discuss a heuristic approach  to include dissipative effects in this description.
	\end{abstract}

	\setcounter{page}{1}


\renewcommand{\thefootnote}{\arabic{footnote}}
\setcounter{footnote}{0}

\section{Introduction}
One of the aims of stochastic inflation is to determine an effective
description for the dynamics of inflationary
fluctuations at superhorizon scales. See e.g.
\cite{Starobinsky:1986fx,Nambu:1987ef,Starobinsky:1994bd,Enqvist:2008kt,Finelli:2008zg}.
Although born as quantum fields at short distances,
a process of classicalization converts small-scale modes into 
classical stochastic variables at scales well larger
than the Hubble horizon \cite{Polarski_1996,KIEFER_1998,Kiefer:2008ku,SUDARSKY_2011,Burgess:2014eoa,Martin_2016,Martin_2022,Chandran_2024}. Long wavelength inflationary
modes sample large super-Hubble regions, in principle assuming different
values in different horizon-size patches. Focusing
on a long-wavelength, coarse-grained  field
controlling scalar fluctuations, 
and assigning to it a probability density, the latter obeys a Fokker-Planck diffusion equation,
with a noise induced by short wavelength modes
in the process of crossing the horizon.
The approach of stochastic inflation is helpful
in dealing with quantum divergences of  light 
scalar fields in de Sitter space, and allows one to 
obtain a full non linear probability  distribution
for the inflationary scalar fluctuations at superhorizon scales (see e.g. \cite{Vennin:2020kng,Green:2022ovz,Burgess:2022rdo}
for recent reviews).

The Fokker-Planck diffusion equation for the field 
probability density can be obtained
from a functional Schr\"odinger formulation \cite{Guth:1985ya,Guven:1987bx},
see  \cite{Burgess:2014eoa,Burgess:2015ajz}.
Here we propose a complementary viewpoint based on
Madelung approach to Schr\"odinger equation \cite{Madelung:1927ksh},
accompanied by an appropriate coarse-graining procedure. 
We find that the system obeys coarse-grained
equations corresponding to the ones
of a superfluid, including
a contribution of quantum
pressure
to the Euler equation. The irrotational
fluid velocity is proportional to the gradient of the wavefunction phase.
The fluid propagates through an abstract coarse-grained scalar space, with
density and velocity 
depending on amplitude and phase
of wavefunction. 
Short wavelength modes, integrated
out from the description, produce
an external
force acting on the fluid. See sections \ref{sec_setp} and \ref{sec_cgd}. The advantage of this perspective is the alternative viewpoint
on the physics involved, which can 
shed new light on known results, and can indicate avenues for possible generalizations.

For  slow-roll inflation
the 
process of classicalisation is realized in terms
of `decoherence without decoherence' \cite{Polarski_1996}, since
the contribution of quantum pressure to the Euler
equation is very rapidly damped by the universe
expansion. Interestingly, this phenomenon does
not occur in an ultra--slow-roll regime, where 
the quantum pressure term is important and contribute
to the fluid dynamics. 
Our approach then provides a novel perspective
on the quantum-to-classical transition during inflation. 
 Our superfluid framework aligns with the conventional Starobinsky formulation of stochastic inflation, while also delivering complementary perspectives -- most notably offering  insight into the phase dynamics of the inflationary wavefunction. See sections \ref{sec_exmp} and \ref{sec_phin}.  We also propose a heuristic approach to our framework, aimed to include
 effects of dissipation in this description. See
 section \ref{sec_hea}.

\section{Set-up}
\label{sec_setp}

We make use of a Schr\"odinger functional approach to analyse the dynamics of scalar fluctuations
during cosmological inflation \cite{Guth:1985ya,Guven:1987bx}. We find it convenient
for formulating a stochastic approach to the system,
and for addressing  from
a novel viewpoint the process
of quantum-to-classical transition in inflationary cosmology.

\smallskip

Our starting point is the quadratic action for free scalar Fourier
modes in quasi-de Sitter space (we assume $\varphi_{-k}\,=\,\varphi_k^*$)
\be
\label{inqsa}
S\,=\,\frac12 \int\, d^3 k\, d\tau\,  \
{z^2(\tau)}\left( \varphi'_k\,\varphi'_{-k}-k^2 \varphi_k \varphi_{-k}\right)\,.
\ee
The function $z(\tau)$, known as the pump field, is model-dependent, and it  characterizes the dynamics of the fluctuations under consideration. We focus on a free
massless field for simplicity, being it 
sufficient for our purposes. 

The quadratic expression \eqref{inqsa} has the generic structure of an action for free fluctuations in single-field inflation. It  describes the evolution of the Mukhanov-Sasaki variable $\zeta_k$, which governs curvature fluctuations in standard slow-roll inflationary models, where $z \propto a(\tau)$
  with $a(\tau)$  the scale factor expressed in terms of the conformal time. Action \eqref{inqsa} can also describe massless spin-2 (tensor) or spin-0 (scalar) fluctuations in pure de Sitter space, again with $z\propto a$. More general scenarios are possible, and warrant further exploration. Interestingly, also ultra-slow roll inflation \cite{Kinney:2005vj,Leach:2000yw,Leach:2001zf} can be studied starting from an action \eqref{inqsa}:
  in this case, a violation of slow-roll conditions leads to a rapid decrease
  of the pump field, $z\propto 1/a^2$. This
scenario is important in models producing primordial black holes. It is interesting
to analyse it from a new perspective, in a framework of stochastic inflation. 
  In the examples that follow, we plan to study
  both the case of SR ($z\propto a$) and USR  ($z\propto 1/a^2$) 
  evolution. 

  \smallskip
  
The  equation of motion for $\varphi_k$ obtained
from eq \eqref{inqsa} results
\be
\label{eq_clp}
\frac{1}{z(\tau)}\partial_\tau^2 \left[ z(\tau)\,\varphi_k \right]+\left( k^2-\frac{z''}{z}\right)\,\varphi_k\,=\,0
\,.
\ee
We assume that the scalar fluctuations
$\varphi_k$ satisy a Wronskian normalization
condition $\varphi'_k \varphi_{-k}-\varphi_k \varphi'_{-k}=i/z^2$ for $k\neq0$, as well as Bunch Davies
conditions at $\tau\to-\infty$. 
This requirement is motivated by the 
underlying quantum behavior of fluctuations at very small scales. The zero mode $\varphi_0$
requires a special treatment. 
The Lagrangian
density ${\cal L}_k$ in Fourier space is the integrand
of action \eqref{inqsa}.  
The 
corresponding conjugate  momentum is 
\be
\pi_{k}\,=\,\frac{\delta {\cal L}_k}{\delta \varphi'_{k}}
\,=\,z^2(\tau)\, \varphi'_{-k}\,,
\ee
which allows us to build the quadratic Hamiltonian for the system:
\be
{\cal H}_k\,=\,\frac{\pi_{k} \pi_{-k}}{z^2(\tau)}
+z^2(\tau) \varphi_{k} \varphi_{-k}\,.
\ee
The functional Schr\"odinger picture  promotes the
fields $\varphi_k$, $\pi_k$ to operators,  equipping them with a hat. A quantum mechanical wave-function $\Psi_k(\varphi_k, \tau)$ 
is introduced, depending on the c-number quantity 
$\varphi_k$ evaluated at conformal time $\tau$. The operators 
 $\hat \varphi_k$, $\hat \pi_k$ act on the wavefunction as
 \bea
 \hat{\varphi}_k\,\Psi_k &=&{\varphi}_k\,\Psi_k\,,
\\
\hat{\pi}_k\,\Psi_k&=&\frac{\hbar}{i}
\frac{\partial\,\Psi_k}{\partial \varphi_k}\,.
 \eea
 Such rules allow us to express
 the corresponding Schr\"odinger equation
 \be
 \label{eq_schr}
i\,\hbar\,\frac{\partial \Psi_k}{\partial \tau}
\,=\,{\cal H}_k\, \Psi_k\,,
\ee
with Hamiltonian
\bea
 \label{eq_qham}
{\cal H}_k&=&
-\frac{\hbar^2}{z^2(\tau)}
\,\frac{\partial^2}{\partial \varphi_k\,\partial \varphi_{-k}}
+z^2(\tau)\,
k^2
 \, \varphi_k\, \varphi_{-k}\,.
\eea
Eqs \eqref{eq_schr}, \eqref{eq_qham} are the starting
point of our treatment. 
  Following Madelung \cite{Madelung:1927ksh,Feynman:1494701} (see \cite{Widrow:1993qq,Uhlemann:2014npa,Hui:2016ltb,Garny:2019noq,Hui:2021tkt,Ferreira:2020fam} for applications of this approach
  to large scale structures and  dark matter scenarios), we decompose the wavefunction $\Psi_k$ in an amplitude and a phase
\be
\label{ans_mad}
\psi_k\,=\,\sqrt{\rho_k}\,e^{i\,z^2(\tau)\,\theta_k/\hbar}
\,.
\ee
Assuming that $\rho_k$ and $\theta_k$ are real
functions of $\varphi_k$ and $\tau$, we can plug Ansatz
\eqref{ans_mad} in the  Schr\"odinger equation \eqref{eq_schr}. Its real
and
imaginary parts lead to a system
of two coupled equations 
\bea
\label{eqone}
0&=&
\frac{\partial \rho_k}{\partial \tau}+ {2 \rho_k} \frac{\partial^2\,\theta_k}{\partial  \varphi_k\, \partial \varphi_{-k}}
+\frac{\partial \rho_k}{\partial \varphi_k}\frac{\partial \theta_k}{\partial \varphi_{-k}}
+\frac{\partial \rho_{k}}{\partial \varphi_{-k}}\frac{\partial \theta_k}{\partial \varphi_{k}}\,,
\\
0&=&
\frac{\partial_\tau \left(z^2  \theta_k \right)}{ z^2(\tau)}+
k^2
\, \varphi_k\, \varphi_{-k}
+\frac{1}{4\, z^4(\tau)\,\rho_k^2}\,\frac{\partial \rho_k}{\partial \varphi_k}\frac{\partial \rho_k}{\partial \varphi_{-k}}
+\frac{\partial \theta_k}{\partial \varphi_k}\frac{\partial \theta_k}{\partial \varphi_{-k}}
-\frac{\hbar^2}{2\, z^4(\tau)\,\rho_k } \frac{\partial^2\, \rho_k}{\partial \varphi_k\,\partial \varphi_{-k}}\,,
\nonumber
\\
\label{eqtwo}
\eea
which resemble the continuity 
and Euler equations of fluid dynamics. One of our aims
 is to follow the evolution of the wavefunction phase, and study its
consequences for the system.  
 An appropriate coarse-graining procedure,
which we discuss next, allows us to combine
equations \eqref{eqone}
and \eqref{eqtwo} in a way
that makes more manifest the connection
with fluid dynamics at superhorizon scales,
and clarify the nature of external forces
acting on the fluid. 

\section{Coarse-grained equations}
\label{sec_cgd}

The dynamics 
of superhorizon quantities is determined by a set of stochastic equations, obtained by a coarse-graining procedure aimed at marginalising over sub-horizon modes. In  section \ref{sec_cgp} and \ref{sec_con} we develop a coarse-grained version
of eqs \eqref{eqone} and \eqref{eqtwo}, showing that they reduce to the equations governing a superfluid. 
After analysing specific applications in section \ref{sec_exmp},
in section \ref{sec_phin} we further discuss    physical consequences of our findings.  

\subsection{ Coarse-graining procedure}
\label{sec_cgp}

We coarse-grain marginalising over
sub-Hubble modes,  focusing  only on large-scale, super-Hubble fields \cite{Wands:2000dp,Salopek:1990jq,Sasaki:1995aw}. In fact, the subhorizon modes do not directly couple
to the superhorizon ones: the former contribute
to the dynamics of the latter only through their
effects at horizon crossing. For complementary perspectives to 
stochastic inflation, see also \cite{Rigopoulos:2003ak,Tanaka:2007gh,Tolley:2008na,Agon:2014uxa,Grain:2017dqa,Cespedes:2023aal,Launay:2024qsm}.

We formally introduce 
coarse-grained quantities 
\bea
\bar \rho &=&\Pi_{k}\,\rho_k\,,
\\
\bar \theta &=& \sum_{k}\,\theta_k\,,
\eea
where the product and the sum are limited
to Fourier modes $k\le a H$. Correspondingly,
the superhorizon wavefunction for the system
is $\bar \Psi=\Pi_k\,\psi_k$.
Physically, 
we identify $\bar \rho$
and $\bar \theta$ as the  fluid energy
density and the velocity potential
in the space of scalar field configurations. (We will later discuss  how the velocity potential, related
with the wavefunction phase,
is  connected to the fluid velocity.)
The coarse-grained scalar is
a real quantity, obtained by summing over Fourier modes,
\be
\bar \varphi({\bf x})\,=\,\frac{1}{\sqrt{2}}\sum_{| {\bf k}|<a H}\,\left(\varphi_k e^{i {\bf k} \cdot {\bf x}}+\varphi_{-k}e^{-i {\bf k} \cdot {\bf x}}\right)
\,.
\ee
The corresponding gradient along the scalar
field direction is defined as
\be
\label{def_nab}
{\bf \nabla}\,\equiv\,
\frac{\partial}{\partial \bar \varphi}\,=\,
\frac{1}{\sqrt{2}}\,\sum_{| {\bf k}|<a H}\,\left(
e^{-i {\bf k} \cdot {\bf x}}\,\frac{\partial}{\partial  \varphi_{k}}
+
e^{i {\bf k} \cdot {\bf x}}\,\frac{\partial}{\partial  \varphi_{-k}}
\right)\,.
\ee

\smallskip
When applying the previous
coarse-graining
definitions to the evolution equations \eqref{eqone} and \eqref{eqtwo}, 
we  further integrate over
space, so to track the relevant contributions which are approximately 
constant over a particular
Hubble volume, ${\cal V}\sim H^{-3}$, centered
say at position $\vec x=0$. This procedure
leads to  simplifications. For example,
\bea
\nonumber
 \vec \nabla^2 \theta&=&\frac{1}{2}\int_{\cal V}\,d^3 x\,
 \left[\sum_{q}\,\left(
e^{-i q x}\,\frac{\partial}{\partial  \varphi_{q}}
+
e^{i q x}\,\frac{\partial}{\partial  \varphi_{-q}}
\right)\right]
 \,\left[ \sum_{k}\,\left(
e^{-i k x}\,\frac{\partial \bar \theta}{\partial  \varphi_{k}}
+
e^{i k x}\,\frac{\partial \bar \theta}{\partial  \varphi_{-k}}
\right)\right]\,,
\\
&=& 
\left(\frac{\partial}{\partial  \varphi_k}\,
\frac{\partial\,\theta_k}{   \partial \varphi_{-k}} 
+\frac{\partial}{\partial  \varphi_{-k}}\,
\frac{\partial\,\theta_k}{   \partial \varphi_{k}}
\right)\,.
\label{eq_nablsq}
 \eea
Hence the spatial
integration allows us to neglect rapidly oscillating pieces which average
to zero. Analogously, 
\be
\nabla \bar \theta \cdot \nabla \bar
\rho\,=\,\frac{\partial  \bar \rho}{\partial \varphi_k}\frac{\partial \theta_k}{\partial \varphi_{-k}}
+\frac{\partial \bar \rho}{\partial \varphi_{-k}}\frac{\partial \theta_k}{\partial \varphi_{k}}
\,.
\ee
In what follows, we also define $\bar \varphi^2 \equiv \sum_k |\varphi_k|^2$. We emphasize that
we use
the gradient symbol ${\bf \nabla}$ to indicate
derivatives along the coarse-grained
field $\bar \varphi$, see eq \eqref{def_nab}.
 
\subsection{ Continuity and Euler equations }
\label{sec_con}

We focus on eq \eqref{eqone} for a given mode $k$, 
and we 
multiply it by the factors $\rho_{k-2}$, $\rho_{k-1}$, $\rho_{k+1}$ etc.
We get
\bea
0&=&\left(
\dots \rho_{k-1}\,\frac{\partial \rho_k}{\partial \tau}\,\rho_{k+1}\dots\right)+ { \bar \rho}\,
\left(\frac{\partial}{\partial  \varphi_k}\,
\frac{\partial\,\theta_k}{   \partial \varphi_{-k}} 
+\frac{\partial}{\partial  \varphi_{-k}}\,
\frac{\partial\,\theta_k}{   \partial \varphi_{k}}
\right)
+\frac{\partial  \bar \rho}{\partial \varphi_k}\frac{\partial \theta_k}{\partial \varphi_{-k}}
+\frac{\partial \bar \rho}{\partial \varphi_{-k}}\frac{\partial \theta_k}{\partial \varphi_{k}}
\,.
\nonumber
\\
\eea
We  substitute  ${\partial \theta_k}/{\partial \varphi_{k}}={\partial \bar \theta}/{\partial \varphi_{k}}$ in the previous expression, since  $\theta_k$
depends on $\varphi_{k}$ only. 
We sum over momentum modes $k$, and we integrate over  a volume ${\cal V}$, using
the definition of coarse-grained quantities, and relations as eq \eqref{eq_nablsq}. We obtain
the expected structure of a fluid continuity equation
\be
\label{eq_cont}
\frac{\partial \bar \rho}{\partial \tau}+\bar \rho\,  \nabla^2  \bar \theta+  \nabla \bar \rho \cdot  \nabla \bar \theta\,=\,0
\,.
\ee
In writing this equation, we assume that the coarse grained quantities $\bar \rho$
and $\bar {\theta}$  depend on conformal time $\tau$, and on the
coarse-grained scalar $\bar \varphi$. The latter
plays the role of spatial coordinate along which the fluid propagates. 

\medskip

A bit more work is needed to obtain an equation
which resembles Euler's. We re-assemble eq \eqref{eqtwo} as
\bea
0&=&\frac{1}{z^2}\frac{\partial (z^2  \theta_k)}{\partial \tau}+k^2\, \varphi_k\, \varphi_{-k}
-\frac{1}{4\, z^4 }\,\frac{\partial (\ln \rho_k)}{\partial \varphi_k}\frac{\partial  (\ln \rho_k)}{\partial \varphi_{-k}}
+\frac{\partial \theta_k}{\partial \varphi_k}\frac{\partial \theta_k}{\partial \varphi_{-k}}
-\frac{\hbar^2}{2\, z^4 }
\frac{\partial}{ \partial \varphi_k}\,\left(
 \frac{\partial   (\ln \rho_k)}{\partial \varphi_{-k}}
\right)\,.
\nonumber
\\
\label{eqtwoA}
\eea
Since each $ \rho_k$, $\theta_k$ depend on the 
a single-mode $
\varphi_k$ only, we  directly substitute the  bar quantities $\bar \rho$ and $\bar \theta$ in all
the terms of eq \eqref{eqtwoA} containing
derivatives along $\varphi_k$. 
We sum over $k$, and we integrate over a volume ${\cal V}$. We obtain a relation 
which corresponds to Euler equation in fluid dynamics
expressed in terms of velocity potential $\bar \theta$:
\bea
0&=&\frac{1}{z^2 }\frac{\partial (z^2  \bar \theta)}{\partial \tau}+\left( \sum_k k^2\, \varphi_k\, \varphi_{-k}\right)
+\frac12 \bar \nabla    \bar \theta \cdot \bar \nabla   \bar \theta
- \frac{\hbar^2\,\bar \nabla^2
  \left(  \bar \rho^{1/2}\right)}{2 \,z^4\,\bar \rho^{1/2}}
  \,.
\label{eqtwoBb}
\eea
The sum 
within square parenthesis
is over
momentum modes $k<a H$. In order to discuss the physical consequences of the coarse-grained
equations, we find convenient to pass from conformal time $\tau$ to number of e-folds of expansion, $d n\,=\,a H\, d \tau$ \cite{Salopek:1990re,Vennin:2015hra}. We rescale the velocity potential -- i.e. the 
phase of the wavefunction -- 
as
\be
\bar \theta(n, \bar \Phi)
\,\equiv\,
a(n)\,H\,
\bar \Theta(n, \bar \Phi) \,.
\ee
We
find the two coupled equations
\bea
\label{eqtwoBca}
0&=&\frac{\partial \bar \rho}{\partial n}+{\bar \rho} \, \nabla^2 \bar \Theta 
+ 
\left( \nabla \bar \rho \right)\left( \nabla \bar \Theta \right)\,,
\\
0&=&\frac{\partial \bar \Theta}{\partial n}+
\frac{\bar \Theta}{a H z^2 }\frac{d\,\left(
a H z^2 
\right)
}{d n} -K(n)\,\bar \varphi^2
+\frac{1}{2} \left(\nabla  \bar \Theta  \right)^2
- \frac{\hbar^2\,\nabla^2 
  \bar \rho^{1/2}}{2\,H^2\,a^2\,z^4\, \bar \rho^{1/2}}  
  \,,
  \label{eqtwoBcb}
\eea
with 
\bea
\label{eq_defok}
K(n)&\equiv&
\frac{1}{a^2 H^2}\,\frac{\int_0^{aH} k^2 |\varphi_k|^2\,d^3 k}{\int_0^{aH}  |\varphi_k|^2\,d^3 k}\,.
\eea
In this definition we substitute the sum with an integral, and we 
adopt the convention to integrate over superhorizon modes from horizon exit $k=a H$ to large scales $k\to0$.
The resulting quantity depends on the e-fold number $n$.

The term proportional to $K(n)$ describes a stochastic external force acting on the fluid
velocity potential, caused by the small-scale
modes crossing the horizon. 
Its origin is  analog to
 the `noise' term in Starobinsky description of stochastic inflation (see section \ref{sec_pers}). From
our perspective, the long wavelength modes forming the 
fluid can be interpreted as an open system coupled to the environment of small-scale modes.
 Equations \eqref{eqtwoBca} and \eqref{eqtwoBcb} are written in terms of velocity potential $\theta$: defining
\be
{\bf v} \,\equiv \, {\bar \nabla} \,\bar \Theta
\ee
we can re-express them in a form which is more recognizable in terms of  a fluid dynamics description:
\bea
\label{eqtwoCca}
0&=&\frac{\partial \bar \rho}{\partial n}+{\bar \nabla} \,\cdot 
\left( \bar \rho\,{\bf v}
\right)\,,
\\
0&=&
\frac{\partial \,{\bf v}}{\partial n}
+
\frac{{\bf v}}{a H z^2 }\frac{d\,\left(
a H z^2 
\right)
}{d n} -2\,K\, \bar \varphi
+{\bf v}\cdot  {\bar \nabla} {\bf v}
- \frac{\hbar^2}{2\,H^2\,a^2\,z^4} 
{\bar \nabla}
\left(\frac{
  {\bar \nabla}^2\bar \rho^{1/2}}{\bar \rho^{1/2}}
  \right)
\nonumber\\
\label{eqtwoCcb}
\eea
These equations describe a (super)fluid flowing in one spatial dimension, represented
by the scalar manifold $\bar \varphi$. 
The second and third terms in the Euler equation 
\eqref{eqtwoCcb} are due to space-time expansion, and to an external force acting on the fluid.
The last contribution proportional to $\hbar^2$ in eq \eqref{eqtwoCcb} 
corresponds to the so-called quantum pressure. As we will see, it plays
a role in systems including phases of ultra--slow-roll. Notice
that eq \eqref{eqtwoCcb} does {\it not} contain the contribution  $(\nabla {\cal P})/\bar \rho$ depending on the internal fluid
pressure ${\cal P}$ \cite{landau1959fm}.
In a sense, our fluid has
no internal `classical' pressure, at least
within the superhorizon framework we are adopting. 
We will return to this point in section \ref{sec_phin}.

\subsection{ The perspective of Starobinsky diffusion equation}
\label{sec_pers}

 There is a relation between our discussion and the usual stochastic
 approach to inflation based
 on the Fokker-Planck equation. The
 square of the wave function $\Psi^* \Psi=\rho$ -- we adopt the Ansatz \eqref{ans_mad} --  can be interpret as the probability  for a  coarse-grained scalar profile to acquire a configuration $\varphi$ at
 superhorizon scales.
 Formally, this quantity is the same thing as the  $\rho$ appearing in eqs \eqref{eqtwoBca} and
\eqref{eqtwoBcb}, but apparently with 
 a distinct physical interpretation. 
  
 Since the scalar system
 we start from in eq \eqref{inqsa} is free, we can assume that such probability $\Psi^* \Psi$ follows a Gaussian distribution, normalised to one upon integration over $\bar \varphi$. Notice that, although the equations governing
 fluctuations are free, nevertheless the coarse-grained
 system has a non-trivial evolution thanks to the noise
 terms induced by short wavelength modes
 crossing the horizon during inflation. This phenomenon leads
 to an open system where long modes interact with the short mode environment. 
 Under our  hypothesis, it is possible to show \cite{Burgess:2014eoa,Burgess:2015ajz,Tasinato:2022asj} (see also Appendix \ref{sec_stanst}) that the free scalar-field   configuration satisfies a diffusion-like, Starobinsky \cite{Starobinsky:1994bd} equation
 \be
 \label{eq_difs}
 \frac{\partial \rho(n, \bar \varphi)}{\partial n}  
 \,=\,\frac{H^2\,{\cal N}(n)}{8 \pi^2}\,
 {\bar \nabla}^2
 \rho(n, \bar \varphi)+{\cal D}(n)\,{\bar \nabla}\left(
 \bar \varphi\,\rho(n, \bar \varphi)\right)\,,
 \ee
with noise  and drift coefficients
are given by a combination of momentum modes
\bea
\label{eq_mno}
{\cal N}(n)&=&\frac{2\,|\varphi_0|^2}{a\,H^3}
\int_{a H}^0\,k^2\,d k\,
\partial_\tau \left(\frac{|\varphi_k|^2}{|\varphi_0|^2} \right)\,,
\\
\label{eq_mdr}
{\cal D}(n)&=&-\frac{1}{2\,a H}
\partial_\tau \ln \left( {|\varphi_0|^2}\right)
\,.
\eea
The stochastic  noise ${\cal N}$ is caused by short modes crossing the horizon during inflation; the drift is driven
by the zero mode $\varphi_0$. 
Eq
\eqref{eq_difs} is complementary to  equations \eqref{eqtwoBca} and
\eqref{eqtwoBcb}: it actually provides  useful
information in dealing with the fluid evolution,
as we will learn through the examples of section \ref{sec_exmp}.

In fact, the three equations \eqref{eqtwoBca}, \eqref{eqtwoBcb}, and \eqref{eq_difs} fully characterize the components
of the wavefunction. 
In short, equations \eqref{eq_difs}, \eqref{eqtwoBca} determine the amplitude $\bar \rho$, 
and the part of the phase $\bar \Theta$ that depends on the position
on the scalar-field space $\bar \varphi$. The Euler equation \eqref{eqtwoBcb}, then, fully determines
the time-dependent part of the phase that does not depend explicitly on $\bar \varphi$.  We will expand on this in section \ref{sec_phin}.

\section{ Examples: slow-roll and ultra slow-roll inflation}
\label{sec_exmp}

We consider  two representative  examples
as applications of the previous results. We focus on evolution in
quasi-de Sitter space, with
scale factor well approximated
by an exponential $a(n)\,\simeq\,e^n$, in
terms of the e-fold number $n$, 
up to small slow-roll corrections which we neglect.
 We are interested in determining the late-time solutions for the fluid energy density $\bar \rho$ 
and velocity potential $\bar \Theta$, neglecting contributions that decay faster than $1/n$ in order to tackle
late-time superhorizon dynamics only. We discuss
two possible inflationary regimes: slow-roll (SR) and ultra--slow-roll (USR) epochs, the latter being relevant
for scenarios leading to  primordial black hole formation.
We will learn that  while the SR evolution
is controlled by classical stochastic equations -- 
thanks to a phenomenon related to decoherence
without decoherence \cite{Polarski_1996} -- the USR equations receive
quantum  
contributions depending on $\hbar$. (See also \cite{Brandenberger:1990bx,Calzetta:1995ys,Lesgourgues:1996jc,Lombardo:2004fr,Burgess:2006jn,Sharman:2007gi,Nelson:2016kjm,Hollowood:2017bil,Burgess:2022nwu,Martin:2018zbe,Colas:2022kfu,DaddiHammou:2022itk,Late-time_decoherence,Lopez:2025arw} for interesting perspectives on
quantum-to-classical transition during inflation.)

\subsection{Slow-roll}

Slow-roll inflation is the leading paradigm for explaining
the initial conditions of the observed universe. In this case we are able to follow
in detail the dynamics of the fluid potential velocity $\bar \Theta$. 
The pump field reads $z(n) = z_0 \,a(n)$,
with $z_0$ a constant depending on the physics we wish to describe. 
The solution
of the mode equation satisfying the requested boundary conditions
is
\be
\varphi_k\,=\,\frac{H}{\sqrt{2}\,k^{3/2}}\,(1+i k \tau)\,
e^{-i k \tau}
\,.
\ee
At sufficiently late times, $n\gg1$,
the coefficient $K$ appearing in eq \eqref{eqtwoBcb} is easily evaluated
using eq \eqref{eq_defok}, resulting
\be
K\,=\,\frac{3}{4\,n}+{\cal O}\left({1}/{n^2} \right)
\,.
\ee
The large $n$ limit implies we focus on late time dynamics
as discussed above. 
 The fluid equations become 
\bea
0&=&\frac{\partial \bar \Theta}{\partial n}+3  \, \bar \Theta -\frac{3}{4\,n}
\,\bar \Phi^2
 +\frac{1}{2 } \left(\frac{ \partial   \bar \Theta}{\partial \bar \varphi}   \right)^2
- \frac{\hbar^2\,e^{-6 n}}{2\, z_0\, \bar \rho^{1/2}}  \frac{\partial ^2
  \left(  \bar \rho^{1/2}\right)}{\partial \bar \varphi^2}
  \,,
\label{eqtwoBcA}
\\
0&=&\frac{\partial \bar \rho}{\partial n}+{\bar \rho} \, \frac{\partial^2 \bar \Theta}{\partial \bar \varphi^2}+
 \frac{\partial \bar \rho}{\partial \bar \varphi}
\, \frac{\partial \bar \Theta}{\partial \bar \varphi}
\,.
\label{eqtwoBcB}
\eea
The contribution of the external force to the Euler equation \eqref{eqtwoBcA} corresponds
to the term $-3 \Bar \Theta^2/(4n)$. It contributes to a 
force on the fluid velocity at superhorizon scales, induced by the short wavelength scalar modes crossing the horizon.
(A contribution associated with a classical fluid pressure  can also be expected (see section \ref{sec_phin}) but it scales
as $1/n^2$,  hence we neglect it in our discussion.)  
The diffusion equation results
\be
 \label{eq_difsA}
 \frac{\partial \bar \rho}{\partial n} 
 \,=\,\frac{H^2}{8 \pi^2}\frac{\partial^2 \bar \rho}{
 \partial \bar \varphi^2
 }
 \,.
 \ee
We can neglect quantum pressure in eq \eqref{eqtwoBcA}, proportional to $\hbar^2$, 
since its contribution is exponentially suppressed in terms of e-fold number. Including it
would be inconsistent in the regime we are interested in. Neglecting
such quantum effects is related with the phenomenon of decoherence
without decoherence, as discussed in \cite{Polarski_1996}. The
rapid expansion of the universe is responsible
for erasing quantum
contributions. 
In this case, then, the system is 
described by classical evolution equations, including stochastic effects associated
with noise (in the diffusion equation) and external force (in the Euler equation). Quantum effects proportional
to $\hbar$ do 
not play a role on the late-time superhorizon evolution of the fluid system.

 The 
 solutions to the previous set of equations, imposing that the fluid density is initially concentrated
 at the origin for $n=0$, result: 
 \bea
\bar \rho (n, \bar \varphi) &=&\frac{\sqrt{2 \pi}}{H\,\sqrt{ n}}
\,\exp{-\frac{2 \pi^2\,\bar \varphi^2}{H^2\,n}
}
\,,
\\
\bar \Theta (n, \bar \varphi)&=&\frac{\bar \varphi^2}{4 n}
\,.
 \eea

The fluid density is described by a Gaussian whose width depends on time, which spreads in the scalar field space. The solution is the same as  
in the usual stochastic formulation to the probability density for the superhorizon scalar field.   The fluid velocity ${\bar v}= \bar \varphi/(2\,n)$ increases in magnitude as we move
away from the origin in  field space, while  its amplitude decreases with time at fixed
position in the scalar field space. Within our approximations,
the velocity potential $\bar \Theta$ has no contributions independent from the fluid  position  $\bar \varphi$. 

\subsection{Ultra--slow-roll}
\label{sec_USR}

Stochastic effects in regime of ultra--slow-roll (USR) evolution have received much attention in the
recent literature -- see e.g. \cite{Pattison:2017mbe,Ezquiaga:2018gbw,Biagetti:2018pjj,Figueroa:2020jkf,Achucarro:2021pdh,Cai:2022erk,Animali:2022otk,Hooshangi:2023kss,Vennin:2024yzl}  -- given their importance for discussing the production
of primordial black holes, see e.g. \cite{Ozsoy:2023ryl} for a review. We discuss this topic within our superfluid
perspective. 
 We express
the pump field as $z=z_0/a^2(n)$, while maintaining
a de Sitter evolution for the scale factor. In 
this case, the amplitude of the would-be decaying mode
actually increases exponentially with the e-fold
number upon crossing the horizon. The role of the nearly constant mode, then, 
is  much suppressed relatively to such would-be decaying mode (which actually increases in size). Let us
see explicitly how our equations describe these phenomena. 
 The fluid equations become
\bea
0&=&\frac{\partial \bar \Theta}{\partial n}-3  \, \bar \Theta -\frac{3}{4\,n}
\,\bar \Phi^2
 +\frac{1}{2 } \left(\frac{ \partial   \bar \Theta}{\partial \bar \varphi}   \right)^2
- \frac{\hbar^2\,e^{6 n}}{2\, z_0\, \bar \rho^{1/2}}  \frac{\partial ^2
  \left(  \bar \rho^{1/2}\right)}{\partial \bar \varphi^2}
  \,,
\label{eqtwoBcA2}
\\
0&=&\frac{\partial \bar \rho}{\partial n}+{\bar \rho} \, \frac{\partial^2 \bar \Theta}{\partial \bar \varphi^2}+
 \frac{\partial \bar \rho}{\partial \bar \varphi}
\, \frac{\partial \bar \Theta}{\partial \bar \varphi}
\,.
\label{eqtwoBcB2}
\eea
Importantly, notice that   the contribution of quantum pressure, proportional
to $\hbar^2$, increases exponentially  with the e-fold
number in
eq \eqref{eqtwoBcB2}. 
The diffusion equation results
\be
 \label{eq_difsA2}
 \frac{\partial \bar \rho}{\partial n} 
 \,=\,\frac{H^2\,e^{6 n}}{8 \pi^2}\frac{\partial^2 \bar\rho}{
 \partial \bar \varphi^2
 }-3
 \,\frac{\partial \,\left(
 \bar \varphi\,\bar\rho\right)}{
 \partial \bar \varphi 
 }
 \,.
 \ee
 The drift term is due to the contribution of the zero mode (see section \ref{sec_pers}). The noise
 term is exponentially enhanced with respect 
 to the SR case of eq \eqref{eq_difsA}. The solutions to the previous set of equations, describing
 a fluid system with energy density localized at the origin for $n=0$, result
\bea
\bar \rho(n, \varphi)&=&\frac{\sqrt{2 \pi}}{H\,e^{3n}\,\sqrt{ n}}
\,\exp{-\frac{2 \pi^2\,\bar \varphi^2}{H^2\,e^{6n}\,n}
}
\,,
\\
\bar \Theta(n,\varphi)
&=&\frac{\left(1+6n \right)\,\bar \varphi^2}{4 n}
+\frac{\hbar^2\,\pi^2}{3\,H^2\,z_0}\,\frac{1}{n}
\,,
\eea
up to corrections that decrease faster than $1/n$. The fluid density is again described by a Gaussian, whose
width is exponentially enhanced in terms of e-fold number. The fluid velocity potential $\bar \Theta$
again depends
on the position in field space. Interestingly,  it also receives a position-independent contribution depending
on $\hbar^2$, associated with
 the quantum pressure in the Euler equation \eqref{eqtwoBcB2}. Hence, in this context even in a late-time limit, 
 quantum effects play a role  for 
 determining the super-horizon evolution of the phase of the waveform. 
  By making use of the approach developed
in \cite{Tasinato:2020vdk,Tasinato:2023ukp}, we also studied the case of a very brief phase of USR sandwiched between two long phases
of slow-roll, without finding qualitative differences with respect to our discussion above.

\smallskip
To conclude, these two examples demonstrate that even free, Gaussian open systems can have a rich interesting dynamics
from the viewpoint of the superfluid equations we derived -- 
thanks to effects of the environment constituted by the modes
crossing the horizon. 

\section{Physical implications}
\label{sec_phin}

After having discussed the coarse-grained equations and their solutions in
representative  cases, in this section we analyse more general physical implications
of our approach, and of the information we gain about the wavefunction phase.

\smallskip

We interpret the continuity and Euler equations \eqref{eqtwoBca} and
\eqref{eqtwoBcb} as describing the dynamics of a pressureless (super)fluid flowing along the single dimension corresponding
to the coarse grained scalar $\bar \varphi$. The fluid flows over super-horizon patches 
with irrotational velocity of size $|{\bf v}|\,=\,|\nabla \bar \Theta|$.  
In this perspective for stochastic inflation
the solution of the fluid density $\rho(n, \varphi)$ is normalised to one, once integrated over the $\bar \varphi$ coordinate. The Euler equation  \eqref{eqtwoBcb} for the velocity potential includes a  term, $-K(n) \bar \varphi^2$, which accounts for the effects of short-wavelength modes over which we marginalize. It  acts
as external  stochastic force for the fluid system. The coefficient $K(n)$ can be explicitly computed from the classical solution of Eq. \eqref{eq_clp} (see Section \ref{sec_exmp} for examples). Importantly, eq \eqref{eqtwoBcb} also contains a  `quantum pressure' contribution
through its last term. Its
role is relevant in certain contexts as in ultra--slow-roll, see section \ref{sec_USR}. In such example, in fact, the solutions of coarse-grained equations contain explicitly a quantum contribution proportional to $\hbar^2$,  affecting
the velocity potential. Our approach can then help
in characterizing the quantum-to-classical transition of
fluctuations during inflation. 

\smallskip

The three equations \eqref{eqtwoBca}, \eqref{eqtwoBcb}, \eqref{eq_difs} can be solved together, fully determining a solution
for the fluid energy density and velocity potential. The solution of the equations for the system allows us to track in detail
all the components of the  inflationary wavefunction discussed in section \ref{sec_cgd}. Not only its amplitude, but also its phase. In the
remaining part of this section, we discuss physical instances where the phase of the wavefunction
is important for understanding the physics involved. 

\smallskip
 We work with dimensionless variables.  We rescale
the coarse-grained value of the superhorizon scalar $\bar \varphi$ through a quantity $x$ as
\be
\bar \varphi\,\equiv\,\frac{H\,x}{2\pi}\,.
\ee
We assume a normalized Gaussian Ansatz for the solution of the fluid density (appropriate
for our free system \eqref{inqsa})
\be
\label{eq_gar}
\rho(n,x)\,=\,\frac{1}{\sqrt{\pi g(n)}}\,e^{-x^2/g(n)}
\,.
\ee
The diffusion equation \eqref{eq_difs} implies that the function $g(n)$ of eq \eqref{eq_gar} satisfies the following equation
\be
g'(n)-4 {\cal N}(n)+2 {\cal D}(n)\, g(n) =0\,,
\ee
whose solution  determines the time-dependent evolution of the Gaussian width $g(n)$ in terms
of noise and drift functions. The continuity equation \eqref{eqtwoBca} then requires that the velocity potential
$\Theta$ assumes the form
\be
\label{eq_gsth}
\Theta(n,x)\,=\,\left[\,\frac{{\cal N}(n)}{g(n)}
-\frac{{\cal D}(n)}{2}
\right]\,x^2
+c(n)\,,
\ee
with a function $c(n)$ finally determined by  Euler equation \eqref{eqtwoBcb}. The quantity $\Theta(n,x)$ is
proportional to 
 the phase of the wavefunction,  as we learned in section \ref{sec_cgd}. 

\smallskip

These general results can be assembled and used  for reconsidering the coarse-grained wavefunction of the system. Using equations \eqref{eq_gar}, \eqref{eq_gsth}, we find that its structure can be expressed as 
(from now on, set $\hbar=1$):
\be
\label{eq_rwf}
\bar \Psi(n,x)\,=\, \frac{1}{\left( \pi g\right)^{1/4}}\,e^{-
\left[1+i\, H\, a\,z^2\,\left(2{\cal N}-{\cal D} g \right) \right]
\,{x^2}/({2 g})+i  H\, a\,z^2\,c}\,.
\ee

\smallskip
All functions of $n$ entering the expression \eqref{eq_rwf} are determined by the equations
of the system, and their solutions.
The phase of the wavefunction, in particular,
controls the off-diagonal terms of the density matrix associated with this system. The complete
wavefunction is  useful for obtaining a distribution in phase space of the fluid elements, which
can be used to study its properties and how they depend on the wavefunction phase. We do so here,   focusing on the physical implications of the
Wigner distribution \cite{Hillery:1983ms,Polarski_1996}
  which allows us to describe the system in a statistical sense.
See also \cite{Habib:1992ci}. 

\subsection{Statistical description}

 Denoting with $p$ the momentum conjugate to the `position' $x$ of the fluid element in scalar field space (see eq \eqref{sec_cgd}), 
the Wigner function gives a normalized,  Gaussian phase-space distribution $f(x,p)$: 
\bea
f(n,x,p)&\equiv&
\int \frac{d	r}{\pi}\,\Psi^* (x+r,n)\,\Psi(x-r, n)\,e^{2 i r p}\,,
\nonumber
\\
&=&\frac{1}{\pi}\,e^{-
\frac{x^2+\left[p g+H a x \left({\cal D} g-2 {\cal N} \right)z^2\right]^2}{g}}
\,,
\label{eq_wig1}
\eea
which can be interpreted as describing the distribution of position and momenta of fluid
elements at superhorizon scales. (For simplicity, we assume the Hubble parameter to be
constant.) Notice that the phase part  of the wavefunction \eqref{eq_rwf} controls the coupling
between the variable $x$ and its conjugate momentum $p$  
in the
exponent of eq \eqref{eq_wig1}.

Then, the fluid system is characterized  by the position $x$ of the fluid element,   and its conjugate momentum $p$ in phase space.  
  The coarse-grained fluid density $\rho(n,x)$ 
  gives the marginal
  probability 
  to find a fluid element at position $x$: it 
  is obtained
 by integrating  over conjugate momenta
\bea
\label{eq_drh}
\rho(n,x)&\equiv&\int dp  f(n,x,p)
\nonumber
\\
&=&\frac{1}{\sqrt{\pi g(n)}}\,e^{-\frac{x^2}{g(n)}}\,,
\eea
matching the results of eq \eqref{eq_gar}. The normalized distribution of conjugate momenta
is obtained integrating over the coordinate $x$
\bea
\label{eq_dqu}
q(n,p)\,\equiv\,\int dx  f(n,x,p)\,=\,
\frac{\sqrt{g} \,\exp \left(-\frac{p^2 g }{ H^2 a^2 z^4 ({\cal D} g-2 {\cal N})^2+1}\right)}{\sqrt{\pi } \sqrt{{ {H}^2 a^2 z^4 ({
\cal D} g-2 {\cal N})^2+1}}}\,.
\eea
 Using the distribution $f(x,p)$ we can also compute the fluid pressure
${\cal P}$, as
\bea
{\cal P}(n,x)&\equiv&\rho(n,x) \left[\int f(n,x,p)\,p^2\,dp-\left(\int f(n,x,p)\,p\,dp \right)^2  \right]\,,
\nonumber
\\
&=&\frac{\rho(n,x)}{2 g(n)}\,,
\label{rel_forp}
\eea
hence the pressure is proportional to the energy density, times a time-dependent factor.
We might expect
that the fluid pressure above contributes to the Euler equation \eqref{eqtwoBcb} through a contribution scaling as
$\nabla {\cal P}/\rho$. This  latter
quantity decreases proportionally to
$\propto 1/g^2(n)$, accordingly to eq
\eqref{rel_forp}. But notice that such term typically gives a subleading contribution
which can be neglected in solving our equations~\footnote{In fact, for the examples
of SR and USR discussed in section \ref{sec_exmp}, such pressure term scale at least as $1/n^2$
and is neglected in the considerations of those systems where we focus
on contributions scaling at most as $1/n$ at late times.}:
this can explain why we did not find the classical pressure contributions
in our analysis.

Defining the quantity
\bea
\sigma_{x^n p^q }&=&\int x^n p^q \,f(n,x,p)\,d x dp\,,
\eea
we find that
the means $\sigma_x$ and $\sigma_p$ associated with the distribution $f(x,p)$ vanish. The variances and the covariance
instead read
\bea
\sigma_{x^2}&=&\frac{g(n)}{2}\,,
\\
\sigma_{p^2}&=&
\left[\frac{1}{2g(n)}+
\frac{{H^2\,a^2(n)\,z^4(n)}
\left({\cal D}(n) g(n)-2 {\cal N}(n)\right)^2}{2g(n)}
\right]\,,
\\
\sigma_{xp}&=&-\frac{H\,a(n)\,z^2(n)}{2}
\left[{\cal D}(n) g(n)-2 {\cal N}(n)\right]\,.
\eea
The covariance $\sigma_{xp}$ is entirely controlled by the phase of the wavefunction. 
Heisenberg uncertainty relation 
can be expressed as
\be
\sqrt{\sigma_{x^2} \sigma_{p^2} }\,=\,\frac12 \sqrt{1+\sigma_{xp}^2}\ge\frac12\,.
\ee
The covariance contribution allows us to
satisfy the previous inequality in a strict sense.  For the examples of section \ref{sec_exmp}, 
we parameterize the pump field $z(n)\,=\,z_0 a^{1-3 c_0}(n)$, with $z_0$ a constant, and $c_0=0,1$
depending on whether we are in SR or USR phases.
We find 
\be
\label{eq_dsxp}
\sigma_{xp}^2\,=\,\frac{z_0^4 H^6}{16 \pi^4} (1+6 \,c_0\, n )^2e^{6 n} \,,
\ee
hence 
\be
\sqrt{\sigma_{x^2} \sigma_{p^2} }\,=\,\frac12 \left[1+ \frac{z_0^4 H^6}{16 \pi^4} (1+6 \,c_0\, n )^2e^{6 n} \right]^{1/2}\,.
\ee
The right-hand-side of this equation, again depending on the wave-function phase, is an exponentially increasing function scaling as $e^{3n}$. Heisenberg
inequality gets more and more satisfied as time flows,  and  the system rapidly classicalises --  both in the SR \cite{Polarski_1996} and USR phases. We find it particularly
interesting that our approach demonstrates that both cases behave quite similarly for
what concerns the process of classicalization
from a perspective of Wigner statistical distribution. 

\medskip
\noindent
{\bf Entropies:} 
%
%
 As further application of our results, we compute the entropies associated
with the distributions
$\rho(n,x)$, $q(n,p)$, $f(n,x,p)$ of eqs \eqref{eq_drh}, \eqref{eq_dqu}, and \eqref{eq_wig1} in the phase
space $(x,q)$ of position and conjugate momentum. Such entropies control subspaces
of the full phase space $(x,p)$. Hence we can expect entropy inequalities to hold \cite{Serafini:2003ke}. 

The entropy for the distribution $\rho(n,x)$ along 
the coordinate $x$
results (we set  the Boltzmann constant to one)
\be
S_x(n)=- \int dx \rho(n,x)\ln \rho(n,x)\,=\,\frac{1}{2}\left(1+\ln [\pi g(n)]\right)
\,.
\ee
In the examples of section \ref{sec_exmp} the function $g(n)$ increases with time, hence the entropy $S_x(n)$ as well. 
The entropy $S_p(n)$ for the distribution of $q(n,p)$ is
\be
S_p(n)=- \int dp q(n,p)\ln q(n,p)\,=\,\frac{1}{2}\left(1+\ln (\pi/g(n))+\ln \left[1+ \sigma_{xp}^2(n)
 \right]\right)\,.
\ee
Notice that it is an increasing function of
e-fold number
$n$,  thanks to the contribution of the covariance $\sigma_{xp}$ of 
the distribution. 
Instead, the entropy associated to the entire Gaussian distribution $f(n, x,p)$
is constant:
\be
S_{\rm tot}\,=\,-\,\int d x dp\,f(n,x,p) \ln f(n,x,p)\,=\, \left(1+\ln \pi \right)\,.
\ee
Hence, the system satisfies the subadditivity condition
\be
S_x(n)+S_p(n)-S_{\rm tot}(n)\,=\,\ln  \left[1+ \sigma_{xp}^2(n)
 \right]\ge0\,,
\ee
with $ \sigma_{xp}^2$ given in eq \eqref{eq_dsxp}. 
The entropy of the sum of the subsystems  (marginalizing over $x$ or over $p$) is,
as expected, much larger than the entropy of the total system, in the limit of large $n$.
Finally, the so-called mutual entropy controls the mutual information among the subsystems
in coordinates $x$ and $p$. It results 
\be
S_{\rm mut}(n)\,\equiv\,-\int d x dp f(n, x,p) \ln\left( \frac{\rho(n,x) \,q(n,p)}{f(n,x,p)}\right)
\,=\,\frac12\,\ln  \left[1+ \sigma_{xp}^2(n)
 \right]\,.
\ee 
As expected, 
it increases with time thanks to  $\sigma_{xp}^2$, controlled
by the phase of the coarse-grained wave-function.  

\smallskip

Hence
the results of this section
demonstrate the important role of the phase of the wavefunction in characterizing
the physics of superhorizon modes, and how the perspective of fluid dynamics allows
to track its evolution.

\section{A heuristic approach to the interacting system}
\label{sec_hea}

In the previous sections, we derived from
first principles evolution equations for super-horizon coarse grained fields, associated with amplitude and phase of the coarse-grained
wavefunction during inflation. For simplicity, we focused
on a free system, with action given in eq \eqref{inqsa}. The resulting
equations correspond to Euler and continuity equations
of fluid dynamics. Interestingly, the free-field
approach can be applied to SR but also to USR scenarios,
the two cases differing by the behaviour of the pump field $z(\tau)$. It would be very interesting
to being able to compute from
first principles the effects of scalar interactions
in our set-up. This is a difficult task~\footnote{See for example the review \cite{Green:2022ovz} discussing this 
topic in the context of Starobinsky stochastic inflation.} which is left for future work.

\smallskip

But for this final section we discuss an alternative, heuristic method
for capturing  the evolution
of super-horizon fluctuations in an interacting scalar-field set-up.
Our aim is to propose a phenomenological approach based on fluid dynamics,
which explicitly include the effects
of external forces, and
 can then be extended  to include phenomena
as dissipation.  
Also in this
case, we wish to organize the evolution equations
as the ones of a (super)-fluid. The difference with what done above
is that we do not proceed starting from first principles -- as in the  
functional Schr\"odinger approach of  section \ref{sec_setp} -- but from the heuristic  manipulation of coarse-grained stochastic equations, whose structure we assume. 

\smallskip

We consider stochastic, large-scale fluctuations
of a scalar field
during inflation, characterized by self-interactions
controlled by a potential $U(\varphi)$. In this
section, we understand the bars, and we express 
the coarse-grained version of 
superhorizon scalar fluctuations as $\varphi$.
We assume the  density $\rho$ satisfies a  diffusion equation 
\be
\label{eq_difsV2}
 {\partial_n \rho(n,  \varphi)} 
 \,=\,\frac{H^2\,{\cal N}(n)}{8 \pi^2}\,\partial_{\varphi}^2 \rho(n,  \varphi)
 +\frac{1}{3 H^2}\partial_\varphi\left[
 {\left(\partial_\varphi U(\varphi)+3
  H^2 {\cal D}(n) \varphi\right)\,\rho(n,  \varphi)}\right]
  \,,
\ee
with $H$ the constant Hubble parameter during inflation. 
The (assumed) structure of the previous equation, which
should be valid deep in a superhorizon regime, is our
starting point. 
When ${\cal N}=1$ and ${\cal D}=0$
we obtain the standard Starobinsky equation  for stochastic evolution in de Sitter space. More
general noise ${\cal N}$ and drift ${\cal D}$
contributions allow us to catch in principle 
deviations from slow-roll evolution, as the USR
phase. (Recall the results of section \ref{sec_exmp}.)

As in the previous sections,
we can interpret  $\rho(n,\varphi)$ in  eq \eqref{eq_difsV2} as the fluid energy density, and 
again we assume
it satisfies a continuity eq as \eqref{eqtwoBca1}
\bea
\label{eqtwoBca1}
\frac{\partial  \rho}{\partial n}+{ \rho} \, \partial_{\varphi}^2  \Theta 
+ 
\left( \partial_{\varphi}  \rho \right)\left( \partial_{\varphi}  \Theta \right)
\,=\,0\,,
\eea
which involves a velocity potential $\Theta(n,\varphi)$,
with ${\bf v}= \partial_{\varphi} \Theta(n,\varphi)$ the
fluid velocity.

Combining eqs \eqref{eq_difsV2} and \eqref{eqtwoBca1}, and integrating along the scalar direction $\varphi$, we find the following expression for the velocity potential
\be
\label{gensolth}
\Theta(n,\varphi)\,=\,-\frac{H^2\,{\cal N}(n)}{8 \pi^2} \,\ln{\left(\frac{\rho(n,\varphi)}{\rho_0(n)}\right)}-\frac{\left[2\,U(\phi)+ 3 H^2 {\cal D}(n)\,\varphi^2 \right]}{6 H^2}\,,
\ee
with ${\rho_0(n)}$ an arbitrary function of $n$. 
Interestingly, 
defining the fluid pressure as
\be
\label{eq_dep}
{\cal P}(n, \varphi)\,=\,\frac{3 H^2}{8 \pi^2} \left({\cal N}+\frac{\partial_n {\cal N}}{3} \right)\,\rho(n, \varphi)\,,
\ee 
we notice that
the velocity potential satisfies an Euler-type equation 
\bea
\label{eq_heeu}
0&=&\partial_n {\Theta}(n,\varphi)+3 {\Theta}(n,\varphi)+\frac12\,\left( \partial_\varphi  {\Theta}(n,\varphi) \right)^2
+\frac{H^4\,{\cal N}^2(n)}{32\,\pi^4}\, \left(\frac{\partial_\varphi^2\,\sqrt{\rho(n,\varphi)}}{\sqrt{\rho(n,\varphi)}}\right)
\nonumber
\\&&+ \int d \varphi \frac{\partial_\varphi{\cal P}(n,\varphi)}{\rho(n,\varphi)}
+W(n,\phi)\,,
\eea
with potential
\bea
W(n,\phi)&=&
\frac{U(\varphi)}{H^2}
-\frac{\left(U'(\varphi)\right)^2}{18 H^4}
+\frac{{\cal N}(n)\,U''(\varphi)}{24 \pi^2}
-\frac{{\cal D}(n)\,\varphi\,U'(\varphi)}{3 H^2}
\nonumber
\\
&&+\left(3 {\cal D}(n)- {\cal D}^2(n)
+{\cal D}'(n)
\right)\frac{\varphi^2}{2}\,,
\eea
representing the effect of external forces. 
At this point,
 some 
 comments are in order:
\begin{itemize}
\item[-] The arguments leading to eqs \eqref{eqtwoBca1} and \eqref{eq_heeu} aims to determine a system
of two coupled equations for the density and velocity  of the fluid. 
After the derivation above has been carried out, we can  
 take these two equations as the {\it fundamental  relations} to solve for determining the system dynamics, without considering  any more  the diffusion-like equation \eqref{eq_difsV2} from which they originate. 
\item[-] The structure of the two equations is different from what found in the previous sections. This being due to the fact that they have not been derived from first principles, but from
a manipulation of more phenomenological expressions, by identifying -- through educated guesses -- the role of each contribution to the Euler equation.
\item[-] 
 Eqs \eqref{eqtwoBca1} and \eqref{eq_heeu}  are classical stochastic equations
independent from quantum effects (there is no $\hbar$).
They aim to be valid at super-horizon scales. The `quantum pressure'--like term of Euler equation proportional
to $H^4$
depends on the noise parameter ${\cal N}$, and the classical pressure ${\cal P}$ is proportional to the energy density
through relation \eqref{eq_dep}. The fluid system feels an external force controlled by the function $W(n,\varphi)$ depending on the scalar potential $U(\varphi)$, 
as well as on the drift term ${\cal D}(n)$.
\end{itemize}
 It is straightforward to show that the equations admit the expected solutions in special cases. In absence of potential, $U(\varphi)=0$, we recover the solutions of section \ref{sec_exmp}. If we turn on the potential, and set ${\cal N}=1$ and ${\cal D}=0$, we find 
the correct equilibrium solution \cite{Starobinsky:1994bd}
\be
\rho(\varphi)\,=\,\rho_0\,e^{-\frac{8 \pi^2 U(\phi)}{3 H^4}}
\,,
\ee
with $\rho_0$ a constant, fixed accordingly to  normalization
as $\rho_0^{-1}= \int_{-\infty}^{\infty} d \varphi \, \exp{
-\frac{8 \pi^2 U(\phi)}{3 H^4}}$, when this integral converges.

\smallskip
What is interesting of this approach is that we can phenomenologically extend it, by adding dissipative contributions
to the Euler equation. We do so using a standard
textbook approach \cite{landau1959fm}.
 In fluid dynamics, dissipation is associated with the viscosity stress tensor, which adds contributions
to the Euler equation depending on second  spatial derivatives acting on the fluid velocity. 
Accordingly,
in our one-dimensional example the (gradient of the) Euler equation
\eqref{eq_heeu} 
 is  expected to become 
\bea
0&=&\partial_n {\bf v}(n,\varphi)+3 {\bf v}(n,\varphi)+{\bf v}(n,\varphi)
\partial_\varphi {\bf v}(n,\varphi)
+\frac{H^4\,{\cal N}^2(n)}{32\,\pi^4}\, \partial_\varphi\,\left(\frac{\partial_\varphi^2\,\sqrt{\rho(n,\varphi)}}{\sqrt{\rho(n,\varphi)}}\right)
\nonumber
\\&&+
\frac{\partial_\varphi{\cal P}(n,\varphi)- \,\partial_\varphi 
\left(
\eta(n,\varphi) \partial_{\varphi}  {\bf v}(n,\varphi)  \right)}{\rho(n,\varphi)}
+\partial_{\varphi}W(n,\phi)\,,
\label{eq_eudi}
\eea
with $\eta(n,\varphi)$ a viscosity coefficient.
This equation can be thought  as a generalization of Navier-Stokes equation in this of fluid in an expanding universe context.  

\smallskip

It would be nice to find quantitative ways to estimate the structure  of $\eta(n,\varphi)$ -- from a purely phenomenological perspectives, or using specific
physical arguments
as fluctuation-dissipation relations. We leave this question open for the time being. But 
if $\eta$ is proportional to the fluid density,
say  $\eta(n,\varphi) = \eta_0(n)\,\rho(n,\varphi) $, 
then the equations
can be solved analytically, at least in the simplest case $U=0$,  ${\cal N}=1$, and ${\cal D}=0$ of evolution in de Sitter space. The solution 
of eqs \eqref{eqtwoBca1} and \eqref{eq_eudi},
at leading order in $1/n$, results
\bea
\rho(n,\varphi)&=&\frac{1}{\sqrt{\pi\,g(n)}}\,e^{-\varphi^2/g(n)}\,,
\\
\Theta(n,\varphi)&=&\frac{g'}{4 g}\,\phi^2\,,
\eea
with
\be
g(n)\,=\,\left(\frac{H^2}{2\pi^2}-\frac{2}{3n}\eta_0(n)\right)\,n\,.
\ee
Hence, the viscous dissipative term can affect the variance of the Gaussian
distribution of the fluid
density, and consequently all the correlation
functions involving fluid elements.

The study of dissipative effects during inflation is an interesting topic, which is currently developed through many fronts, also using an open effective field theory approach \cite{breuerTheoryOpenQuantum2002,Calzetta:2008iqa,Boyanovsky:2015xoa,Boyanovsky:2015tba,Boyanovsky:2015jen,Burgess:2015ajz,Burgess:2020tbq,Burgess:2022rdo,Colas:2023wxa,Salcedo:2024smn,Salcedo:2024nex}. It will be interesting to understand whether our approach based on a fluid description
of superhorizon fluctuations can help to further develop and investigate this subject.

\section{Outlook}

We developed a superfluid approach to describe
the physics of long wavelength fluctuations in 
a framework related with stochastic inflation. 
We did so by making use of the Madelung approach
to the functional Schr\"oding equation for the 
inflationary wavefunction. We shown
that our method allows to consistently control
the evolution of the inflationary wavefunction, 
and discussed physical implications
of our approach. We pointed out that the quantum
pressure characterizing the Euler equation
for the superfluid can have an important role during
an ultra--slow-roll phase of inflationary dynamics.
Hence, our approach can provide an alternative perspective
on the quantum-to-classical transition during inflation. 
By implementing heuristic, phenomenological
considerations, we also proposed how to include
dissipative effects in our description.

\smallskip
Much work remains to be done. An interesting direction would be to extend the analysis to systems with multiple fields and to explore vortex solutions of the fluid equations that exhibit conservation of vorticity. It is also important to incorporate interaction effects and nonlinearities more systematically, possibly by employing a Gross--Pitaevskii version of the Schr\"odinger equation. Including dissipative effects from first principles would be another valuable step. Finally, it would be exciting to investigate whether the ideas developed here can contribute to the design of \emph{analog cosmology} systems: condensed matter experiments that aim to reproduce key features of inflationary dynamics in the laboratory.

\subsection*{Acknowledgments}
It is a pleasure to thank Thomas Colas for a discussion. GT is partially funded by the STFC grants ST/T000813/1 and ST/X000648/1.  For the purpose of open access, the author has applied a Creative Commons Attribution licence to any Author Accepted Manuscript version arising.

\begin{appendix}

\section{ Standard stochastic
formulas}
\label{sec_stanst}

Since ours is a free field, we assume the following
Gaussian Ansatz for the wavefunction of eq \eqref{eq_schr}:
\be
\Psi_k\,=\,\Omega_k(\tau) e^{-z^2(\tau)\left(\alpha_k(\tau)
\varphi_k \varphi_{-k}-\alpha_0(k)\,\delta_{k0} \varphi_k \right) }\,.
\ee
The Schroedinger equation \eqref{eq_schr} imposes
the conditions
\bea
0&=&\Omega'_k+i \alpha_k \Omega_k\,,
\\
0&=&\alpha'_k+i \alpha_k^2+\frac{2\,z'(\tau)}{z(\tau)}\alpha_k\,.
-i k^2
\label{secst2}
\eea
By defining 
\be
\alpha_k(\tau)\,=\,-i\,\partial_\tau\,\ln{\left[{u_{k}^*}
\right]}\,,
\ee
we find that the second condition above is equivalent
to the  equation \eqref{eq_clp} for the mode $\varphi_k$:
\be
(z u_k)''+\left(k^2-\frac{z''}{z} \right)\,z  u_k=0\,.
\ee
We impose the Wronskian normalization $u_k' u_{-k}-u_{-k}' u_k=i/z^2$, and
Bunch-Davies
boundary conditions at early times. Hence, from now on, we identify $u_k=\varphi_k$. 
\bea
\alpha_k+\alpha_{-k}&=&-\frac{1}{z^2\,\varphi_k \varphi_{-k}}\,,
\\
\alpha_k-\alpha_{-k}&=&-i \,\partial_\tau\ln{\left[{\varphi_k \varphi_{-k}}\right]}
\,.
\eea
Defining $\rho_k=\Psi_k^* \Psi_k$, it
satisfies
the relation
\be
\frac{\partial\,\rho_k}{\partial \tau}
\,=\,
\omega_k \,\frac{\partial^2 \rho_k}{
\partial \varphi_k \partial \varphi_{-k}
}+\omega_0 \left[
\frac{\partial}{\partial \varphi_k}
\left( \varphi_k \rho_k \right)
+
\frac{\partial}{\partial \varphi_{-k}}
\left( \varphi_{-k} \rho_{-k} \right)
\right]\,,
\ee
with
\bea
\omega_k&=&\frac{i}{z^2}
\frac{\alpha_k-\alpha_0-\alpha_{k}^*+\alpha_0^*}{\alpha_k+\alpha_k^*}
\,=\,-{|\varphi_0|^2}
\,\partial_\tau \left(\frac{|\varphi_k|^2}{|\varphi_0|^2} \right)\,,
\\
\omega_0&=&-\frac{i}{2} (\alpha_0-\alpha_0^*)
\,=\,-\frac12 \,\partial_\tau \ln \left( {|\varphi_0|^2}\right)\,.
\eea
Coarse graining as above, we
get a Starobinsky diffusion equation
\be
 \frac{\partial \bar \rho}{\partial n} 
 \,=\,{\cal N}\,\frac{\partial^2 \bar \rho}{
 \partial \bar \Phi^2
 }+{\cal D}\frac{\partial \,\left(
 \bar \Phi\,\bar \rho\right)}{
 \partial \bar \Phi 
 }\,,
 \ee
with noise and drift terms given by 
\bea
{\cal N}&=&\frac{|\varphi_0|^2}{4\pi^2\,a\,H}
\int_{a H}^0\,k^2\,d k\,
\partial_\tau \left(\frac{|\varphi_k|^2}{|\varphi_0|^2} \right)\,,
\\
{\cal D}&=&-\frac{1}{2\,a H}
\partial_\tau \ln \left( {|\varphi_0|^2}\right)\,.
\eea
reproducing eqs \eqref{eq_mno} and \eqref{eq_mdr}.

\end{appendix}

\providecommand{\href}[2]{#2}\begingroup\raggedright\endgroup


\begin{thebibliography}{10}

\bibitem{Starobinsky:1986fx}
A.~A. Starobinsky, ``{STOCHASTIC DE SITTER (INFLATIONARY) STAGE IN THE EARLY
  UNIVERSE},'' \href{https://dx.doi.org/10.1007/3-540-16452-9_6}{{\em Lect.
  Notes Phys.} {\bfseries 246} (1986) 107--126}.

\bibitem{Nambu:1987ef}
Y.~Nambu and M.~Sasaki, ``{Stochastic Stage of an Inflationary Universe
  Model},'' \href{https://dx.doi.org/10.1016/0370-2693(88)90974-4}{{\em Phys.
  Lett. B} {\bfseries 205} (1988) 441--446}.

\bibitem{Starobinsky:1994bd}
A.~A. Starobinsky and J.~Yokoyama, ``{Equilibrium state of a selfinteracting
  scalar field in the De Sitter background},''
  \href{https://dx.doi.org/10.1103/PhysRevD.50.6357}{{\em Phys. Rev. D}
  {\bfseries 50} (1994) 6357--6368},
  \href{https://arxiv.org/abs/astro-ph/9407016}{{\ttfamily
  arXiv:astro-ph/9407016}}.

\bibitem{Enqvist:2008kt}
K.~Enqvist, S.~Nurmi, D.~Podolsky, and G.~I. Rigopoulos, ``{On the divergences
  of inflationary superhorizon perturbations},''
  \href{https://dx.doi.org/10.1088/1475-7516/2008/04/025}{{\em JCAP} {\bfseries
  04} (2008) 025}, \href{https://arxiv.org/abs/0802.0395}{{\ttfamily
  arXiv:0802.0395 [astro-ph]}}.

\bibitem{Finelli:2008zg}
F.~Finelli, G.~Marozzi, A.~A. Starobinsky, G.~P. Vacca, and G.~Venturi,
  ``{Generation of fluctuations during inflation: Comparison of stochastic and
  field-theoretic approaches},''
  \href{https://dx.doi.org/10.1103/PhysRevD.79.044007}{{\em Phys. Rev. D}
  {\bfseries 79} (2009) 044007},
  \href{https://arxiv.org/abs/0808.1786}{{\ttfamily arXiv:0808.1786 [hep-th]}}.

\bibitem{Polarski_1996}
D.~Polarski and A.~A. Starobinsky, ``Semiclassicality and decoherence of
  cosmological perturbations,''
  \href{https://dx.doi.org/10.1088/0264-9381/13/3/006}{{\em Classical and
  Quantum Gravity} {\bfseries 13} no.~3, (Mar., 1996) 377–391}.
  \url{http://dx.doi.org/10.1088/0264-9381/13/3/006}.

\bibitem{KIEFER_1998}
C.~Kiefer, D.~Polarski, and A.~A. Starobinsky, ``Quantum-to-classical
  transition for fluctuations in the early universe,''
  \href{https://dx.doi.org/10.1142/s0218271898000292}{{\em International
  Journal of Modern Physics D} {\bfseries 07} no.~03, (June, 1998) 455–462}.
  \url{http://dx.doi.org/10.1142/S0218271898000292}.

\bibitem{Kiefer:2008ku}
C.~Kiefer and D.~Polarski, ``{Why do cosmological perturbations look classical
  to us?},'' \href{https://dx.doi.org/10.1166/asl.2009.1023}{{\em Adv. Sci.
  Lett.} {\bfseries 2} (2009) 164--173},
  \href{https://arxiv.org/abs/0810.0087}{{\ttfamily arXiv:0810.0087
  [astro-ph]}}.

\bibitem{SUDARSKY_2011}
D.~Sudarsky, ``Shortcomings in the understanding of why cosmological
  perturbations look classical,''
  \href{https://dx.doi.org/10.1142/s0218271811018937}{{\em International
  Journal of Modern Physics D} {\bfseries 20} no.~04, (Apr., 2011) 509–552}.
  \url{http://dx.doi.org/10.1142/S0218271811018937}.

\bibitem{Burgess:2014eoa}
C.~P. Burgess, R.~Holman, G.~Tasinato, and M.~Williams, ``{EFT Beyond the
  Horizon: Stochastic Inflation and How Primordial Quantum Fluctuations Go
  Classical},'' \href{https://dx.doi.org/10.1007/JHEP03(2015)090}{{\em JHEP}
  {\bfseries 03} (2015) 090}, \href{https://arxiv.org/abs/1408.5002}{{\ttfamily
  arXiv:1408.5002 [hep-th]}}.

\bibitem{Martin_2016}
J.~Martin and V.~Vennin, ``Quantum discord of cosmic inflation: Can we show
  that cmb anisotropies are of quantum-mechanical origin?''
  \href{https://dx.doi.org/10.1103/physrevd.93.023505}{{\em Physical Review D}
  {\bfseries 93} no.~2, (Jan., 2016) }.
  \url{http://dx.doi.org/10.1103/PhysRevD.93.023505}.

\bibitem{Martin_2022}
J.~Martin, A.~Micheli, and V.~Vennin, ``Discord and decoherence,''
  \href{https://dx.doi.org/10.1088/1475-7516/2022/04/051}{{\em Journal of
  Cosmology and Astroparticle Physics} {\bfseries 2022} no.~04, (Apr., 2022)
  051}. \url{http://dx.doi.org/10.1088/1475-7516/2022/04/051}.

\bibitem{Chandran_2024}
S.~M. Chandran, K.~Rajeev, and S.~Shankaranarayanan, ``Real-space
  quantum-to-classical transition of time dependent background fluctuations,''
  \href{https://dx.doi.org/10.1103/physrevd.109.023503}{{\em Physical Review D}
  {\bfseries 109} no.~2, (Jan., 2024) }.
  \url{http://dx.doi.org/10.1103/PhysRevD.109.023503}.

\bibitem{Vennin:2020kng}
V.~Vennin, {\em {Stochastic inflation and primordial black holes}}.
\newblock PhD thesis, U. Paris-Saclay, 6, 2020.
\newblock \href{https://arxiv.org/abs/2009.08715}{{\ttfamily arXiv:2009.08715
  [astro-ph.CO]}}.

\bibitem{Green:2022ovz}
D.~Green, {\em {EFT for de Sitter Space}}.
\newblock 2023.
\newblock \href{https://arxiv.org/abs/2210.05820}{{\ttfamily arXiv:2210.05820
  [hep-th]}}.

\bibitem{Burgess:2022rdo}
C.~P. Burgess and G.~Kaplanek, ``{Gravity, Horizons and Open EFTs},''
  \href{https://arxiv.org/abs/2212.09157}{{\ttfamily arXiv:2212.09157
  [hep-th]}}.

\bibitem{Guth:1985ya}
A.~H. Guth and S.-Y. Pi, ``{The Quantum Mechanics of the Scalar Field in the
  New Inflationary Universe},''
  \href{https://dx.doi.org/10.1103/PhysRevD.32.1899}{{\em Phys. Rev. D}
  {\bfseries 32} (1985) 1899--1920}.

\bibitem{Guven:1987bx}
J.~Guven, B.~Lieberman, and C.~T. Hill, ``{Schrodinger Picture Field Theory in
  Robertson-walker Flat Space-times},''
  \href{https://dx.doi.org/10.1103/PhysRevD.39.438}{{\em Phys. Rev. D}
  {\bfseries 39} (1989) 438}.

\bibitem{Burgess:2015ajz}
C.~P. Burgess, R.~Holman, and G.~Tasinato, ``{Open EFTs, IR effects
  \textbackslash{}\& late-time resummations: systematic corrections in
  stochastic inflation},''
  \href{https://dx.doi.org/10.1007/JHEP01(2016)153}{{\em JHEP} {\bfseries 01}
  (2016) 153}, \href{https://arxiv.org/abs/1512.00169}{{\ttfamily
  arXiv:1512.00169 [gr-qc]}}.

\bibitem{Madelung:1927ksh}
E.~Madelung, ``{Quantentheorie in hydrodynamischer Form},''
  \href{https://dx.doi.org/10.1007/BF01400372}{{\em Z. Phys.} {\bfseries 40}
  no.~3, (1927) 322--326}.

\bibitem{Kinney:2005vj}
W.~H. Kinney, ``{Horizon crossing and inflation with large eta},''
  \href{https://dx.doi.org/10.1103/PhysRevD.72.023515}{{\em Phys. Rev. D}
  {\bfseries 72} (2005) 023515},
  \href{https://arxiv.org/abs/gr-qc/0503017}{{\ttfamily arXiv:gr-qc/0503017}}.

\bibitem{Leach:2000yw}
S.~M. Leach and A.~R. Liddle, ``{Inflationary perturbations near horizon
  crossing},'' \href{https://dx.doi.org/10.1103/PhysRevD.63.043508}{{\em Phys.
  Rev. D} {\bfseries 63} (2001) 043508},
  \href{https://arxiv.org/abs/astro-ph/0010082}{{\ttfamily
  arXiv:astro-ph/0010082}}.

\bibitem{Leach:2001zf}
S.~M. Leach, M.~Sasaki, D.~Wands, and A.~R. Liddle, ``{Enhancement of
  superhorizon scale inflationary curvature perturbations},''
  \href{https://dx.doi.org/10.1103/PhysRevD.64.023512}{{\em Phys. Rev. D}
  {\bfseries 64} (2001) 023512},
  \href{https://arxiv.org/abs/astro-ph/0101406}{{\ttfamily
  arXiv:astro-ph/0101406}}.

\bibitem{Feynman:1494701}
R.~P. Feynman, R.~B. Leighton, and M.~Sands, {\em {The Feynman lectures on
  physics; New millennium ed.}}
\newblock Basic Books, New York, NY, 2010.
\newblock \url{https://cds.cern.ch/record/1494701}.
\newblock Originally published 1963-1965.

\bibitem{Widrow:1993qq}
L.~M. Widrow and N.~Kaiser, ``{Using the Schrodinger equation to simulate
  collisionless matter},'' {\em Astrophys. J. Lett.} {\bfseries 416} (1993)
  L71--L74.

\bibitem{Uhlemann:2014npa}
C.~Uhlemann, M.~Kopp, and T.~Haugg, ``{Schr\"odinger method as $N$-body double
  and UV completion of dust},''
  \href{https://dx.doi.org/10.1103/PhysRevD.90.023517}{{\em Phys. Rev. D}
  {\bfseries 90} no.~2, (2014) 023517},
  \href{https://arxiv.org/abs/1403.5567}{{\ttfamily arXiv:1403.5567
  [astro-ph.CO]}}.

\bibitem{Hui:2016ltb}
L.~Hui, J.~P. Ostriker, S.~Tremaine, and E.~Witten, ``{Ultralight scalars as
  cosmological dark matter},''
  \href{https://dx.doi.org/10.1103/PhysRevD.95.043541}{{\em Phys. Rev. D}
  {\bfseries 95} no.~4, (2017) 043541},
  \href{https://arxiv.org/abs/1610.08297}{{\ttfamily arXiv:1610.08297
  [astro-ph.CO]}}.

\bibitem{Garny:2019noq}
M.~Garny, T.~Konstandin, and H.~Rubira, ``{The Schr\"odinger-Poisson method for
  Large-Scale Structure},''
  \href{https://dx.doi.org/10.1088/1475-7516/2020/04/003}{{\em JCAP} {\bfseries
  04} (2020) 003}, \href{https://arxiv.org/abs/1911.04505}{{\ttfamily
  arXiv:1911.04505 [astro-ph.CO]}}.

\bibitem{Hui:2021tkt}
L.~Hui, ``{Wave Dark Matter},''
  \href{https://dx.doi.org/10.1146/annurev-astro-120920-010024}{{\em Ann. Rev.
  Astron. Astrophys.} {\bfseries 59} (2021) 247--289},
  \href{https://arxiv.org/abs/2101.11735}{{\ttfamily arXiv:2101.11735
  [astro-ph.CO]}}.

\bibitem{Ferreira:2020fam}
E.~G.~M. Ferreira, ``{Ultra-light dark matter},''
  \href{https://dx.doi.org/10.1007/s00159-021-00135-6}{{\em Astron. Astrophys.
  Rev.} {\bfseries 29} no.~1, (2021) 7},
  \href{https://arxiv.org/abs/2005.03254}{{\ttfamily arXiv:2005.03254
  [astro-ph.CO]}}.

\bibitem{Wands:2000dp}
D.~Wands, K.~A. Malik, D.~H. Lyth, and A.~R. Liddle, ``{A New approach to the
  evolution of cosmological perturbations on large scales},''
  \href{https://dx.doi.org/10.1103/PhysRevD.62.043527}{{\em Phys. Rev. D}
  {\bfseries 62} (2000) 043527},
  \href{https://arxiv.org/abs/astro-ph/0003278}{{\ttfamily
  arXiv:astro-ph/0003278}}.

\bibitem{Salopek:1990jq}
D.~S. Salopek and J.~R. Bond, ``{Nonlinear evolution of long wavelength metric
  fluctuations in inflationary models},''
  \href{https://dx.doi.org/10.1103/PhysRevD.42.3936}{{\em Phys. Rev. D}
  {\bfseries 42} (1990) 3936--3962}.

\bibitem{Sasaki:1995aw}
M.~Sasaki and E.~D. Stewart, ``{A General analytic formula for the spectral
  index of the density perturbations produced during inflation},''
  \href{https://dx.doi.org/10.1143/PTP.95.71}{{\em Prog. Theor. Phys.}
  {\bfseries 95} (1996) 71--78},
  \href{https://arxiv.org/abs/astro-ph/9507001}{{\ttfamily
  arXiv:astro-ph/9507001}}.

\bibitem{Rigopoulos:2003ak}
G.~I. Rigopoulos and E.~P.~S. Shellard, ``{The separate universe approach and
  the evolution of nonlinear superhorizon cosmological perturbations},''
  \href{https://dx.doi.org/10.1103/PhysRevD.68.123518}{{\em Phys. Rev. D}
  {\bfseries 68} (2003) 123518},
  \href{https://arxiv.org/abs/astro-ph/0306620}{{\ttfamily
  arXiv:astro-ph/0306620}}.

\bibitem{Tanaka:2007gh}
Y.~Tanaka and M.~Sasaki, ``{Gradient expansion approach to nonlinear
  superhorizon perturbations. II. A Single scalar field},''
  \href{https://dx.doi.org/10.1143/PTP.118.455}{{\em Prog. Theor. Phys.}
  {\bfseries 118} (2007) 455--473},
  \href{https://arxiv.org/abs/0706.0678}{{\ttfamily arXiv:0706.0678 [gr-qc]}}.

\bibitem{Tolley:2008na}
A.~J. Tolley and M.~Wyman, ``{Stochastic Inflation Revisited: Non-Slow Roll
  Statistics and DBI Inflation},''
  \href{https://dx.doi.org/10.1088/1475-7516/2008/04/028}{{\em JCAP} {\bfseries
  04} (2008) 028}, \href{https://arxiv.org/abs/0801.1854}{{\ttfamily
  arXiv:0801.1854 [hep-th]}}.

\bibitem{Agon:2014uxa}
C.~Agon, V.~Balasubramanian, S.~Kasko, and A.~Lawrence, ``{Coarse Grained
  Quantum Dynamics},''
  \href{https://dx.doi.org/10.1103/PhysRevD.98.025019}{{\em Phys. Rev. D}
  {\bfseries 98} no.~2, (2018) 025019},
  \href{https://arxiv.org/abs/1412.3148}{{\ttfamily arXiv:1412.3148 [hep-th]}}.

\bibitem{Grain:2017dqa}
J.~Grain and V.~Vennin, ``{Stochastic inflation in phase space: Is slow roll a
  stochastic attractor?},''
  \href{https://dx.doi.org/10.1088/1475-7516/2017/05/045}{{\em JCAP} {\bfseries
  05} (2017) 045}, \href{https://arxiv.org/abs/1703.00447}{{\ttfamily
  arXiv:1703.00447 [gr-qc]}}.

\bibitem{Cespedes:2023aal}
S.~C\'espedes, A.-C. Davis, and D.-G. Wang, ``{On the IR divergences in de
  Sitter space: loops, resummation and the semi-classical wavefunction},''
  \href{https://dx.doi.org/10.1007/JHEP04(2024)004}{{\em JHEP} {\bfseries 04}
  (2024) 004}, \href{https://arxiv.org/abs/2311.17990}{{\ttfamily
  arXiv:2311.17990 [hep-th]}}.

\bibitem{Launay:2024qsm}
Y.~L. Launay, G.~I. Rigopoulos, and E.~P.~S. Shellard, ``{Stochastic inflation
  in general relativity},''
  \href{https://dx.doi.org/10.1103/PhysRevD.109.123523}{{\em Phys. Rev. D}
  {\bfseries 109} no.~12, (2024) 123523},
  \href{https://arxiv.org/abs/2401.08530}{{\ttfamily arXiv:2401.08530
  [gr-qc]}}.

\bibitem{Salopek:1990re}
D.~S. Salopek and J.~R. Bond, ``{Stochastic inflation and nonlinear gravity},''
  \href{https://dx.doi.org/10.1103/PhysRevD.43.1005}{{\em Phys. Rev. D}
  {\bfseries 43} (1991) 1005--1031}.

\bibitem{Vennin:2015hra}
V.~Vennin and A.~A. Starobinsky, ``{Correlation Functions in Stochastic
  Inflation},'' \href{https://dx.doi.org/10.1140/epjc/s10052-015-3643-y}{{\em
  Eur. Phys. J. C} {\bfseries 75} (2015) 413},
  \href{https://arxiv.org/abs/1506.04732}{{\ttfamily arXiv:1506.04732
  [hep-th]}}.

\bibitem{Tasinato:2022asj}
G.~Tasinato, ``{Stochastic approach to gravitational waves from inflation},''
  \href{https://dx.doi.org/10.1103/PhysRevD.105.023521}{{\em Phys. Rev. D}
  {\bfseries 105} no.~2, (2022) 023521},
  \href{https://arxiv.org/abs/2201.10333}{{\ttfamily arXiv:2201.10333
  [hep-th]}}.

\bibitem{Brandenberger:1990bx}
R.~H. Brandenberger, R.~Laflamme, and M.~Mijic, ``{Classical Perturbations From
  Decoherence of Quantum Fluctuations in the Inflationary Universe},''
  \href{https://dx.doi.org/10.1142/S0217732390002651}{{\em Mod. Phys. Lett. A}
  {\bfseries 5} (1990) 2311--2318}.

\bibitem{Calzetta:1995ys}
E.~Calzetta and B.~L. Hu, ``{Quantum fluctuations, decoherence of the mean
  field, and structure formation in the early universe},''
  \href{https://dx.doi.org/10.1103/PhysRevD.52.6770}{{\em Phys. Rev. D}
  {\bfseries 52} (1995) 6770--6788},
  \href{https://arxiv.org/abs/gr-qc/9505046}{{\ttfamily arXiv:gr-qc/9505046}}.

\bibitem{Lesgourgues:1996jc}
J.~Lesgourgues, D.~Polarski, and A.~A. Starobinsky, ``{Quantum to classical
  transition of cosmological perturbations for nonvacuum initial states},''
  \href{https://dx.doi.org/10.1016/S0550-3213(97)00224-1}{{\em Nucl. Phys. B}
  {\bfseries 497} (1997) 479--510},
  \href{https://arxiv.org/abs/gr-qc/9611019}{{\ttfamily arXiv:gr-qc/9611019}}.

\bibitem{Lombardo:2004fr}
F.~C. Lombardo, ``{Influence functional approach to decoherence during
  inflation},'' \href{https://dx.doi.org/10.1590/S0103-97332005000300005}{{\em
  Braz. J. Phys.} {\bfseries 35} (2005) 391--396},
  \href{https://arxiv.org/abs/gr-qc/0412069}{{\ttfamily arXiv:gr-qc/0412069}}.

\bibitem{Burgess:2006jn}
C.~P. Burgess, R.~Holman, and D.~Hoover, ``{Decoherence of inflationary
  primordial fluctuations},''
  \href{https://dx.doi.org/10.1103/PhysRevD.77.063534}{{\em Phys. Rev. D}
  {\bfseries 77} (2008) 063534},
  \href{https://arxiv.org/abs/astro-ph/0601646}{{\ttfamily
  arXiv:astro-ph/0601646}}.

\bibitem{Sharman:2007gi}
J.~W. Sharman and G.~D. Moore, ``{Decoherence due to the Horizon after
  Inflation},'' \href{https://dx.doi.org/10.1088/1475-7516/2007/11/020}{{\em
  JCAP} {\bfseries 11} (2007) 020},
  \href{https://arxiv.org/abs/0708.3353}{{\ttfamily arXiv:0708.3353 [gr-qc]}}.

\bibitem{Nelson:2016kjm}
E.~Nelson, ``{Quantum Decoherence During Inflation from Gravitational
  Nonlinearities},''
  \href{https://dx.doi.org/10.1088/1475-7516/2016/03/022}{{\em JCAP} {\bfseries
  03} (2016) 022}, \href{https://arxiv.org/abs/1601.03734}{{\ttfamily
  arXiv:1601.03734 [gr-qc]}}.

\bibitem{Hollowood:2017bil}
T.~J. Hollowood and J.~I. McDonald, ``{Decoherence, discord and the quantum
  master equation for cosmological perturbations},''
  \href{https://dx.doi.org/10.1103/PhysRevD.95.103521}{{\em Phys. Rev. D}
  {\bfseries 95} no.~10, (2017) 103521},
  \href{https://arxiv.org/abs/1701.02235}{{\ttfamily arXiv:1701.02235
  [gr-qc]}}.

\bibitem{Burgess:2022nwu}
C.~P. Burgess, R.~Holman, G.~Kaplanek, J.~Martin, and V.~Vennin, ``{Minimal
  decoherence from inflation},''
  \href{https://dx.doi.org/10.1088/1475-7516/2023/07/022}{{\em JCAP} {\bfseries
  07} (2023) 022}, \href{https://arxiv.org/abs/2211.11046}{{\ttfamily
  arXiv:2211.11046 [hep-th]}}.

\bibitem{Martin:2018zbe}
J.~Martin and V.~Vennin, ``{Observational constraints on quantum decoherence
  during inflation},''
  \href{https://dx.doi.org/10.1088/1475-7516/2018/05/063}{{\em JCAP} {\bfseries
  05} (2018) 063}, \href{https://arxiv.org/abs/1801.09949}{{\ttfamily
  arXiv:1801.09949 [astro-ph.CO]}}.

\bibitem{Colas:2022kfu}
T.~Colas, J.~Grain, and V.~Vennin, ``{Quantum recoherence in the early
  universe},'' \href{https://dx.doi.org/10.1209/0295-5075/acdd94}{{\em EPL}
  {\bfseries 142} no.~6, (2023) 69002},
  \href{https://arxiv.org/abs/2212.09486}{{\ttfamily arXiv:2212.09486
  [gr-qc]}}.

\bibitem{DaddiHammou:2022itk}
A.~Daddi~Hammou and N.~Bartolo, ``{Cosmic decoherence: primordial power spectra
  and non-Gaussianities},''
  \href{https://dx.doi.org/10.1088/1475-7516/2023/04/055}{{\em JCAP} {\bfseries
  04} (2023) 055}, \href{https://arxiv.org/abs/2211.07598}{{\ttfamily
  arXiv:2211.07598 [astro-ph.CO]}}.

\bibitem{Late-time_decoherence}
C.~Burgess, T.~Colas, R.~Holman, G.~Kaplanek, and V.~Vennin, ``Cosmic purity
  lost: perturbative and resummed late-time inflationary decoherence,''
  \href{https://dx.doi.org/10.1088/1475-7516/2024/08/042}{{\em Journal of
  Cosmology and Astroparticle Physics} {\bfseries 2024} no.~08, (Aug., 2024)
  042}. \url{http://dx.doi.org/10.1088/1475-7516/2024/08/042}.

\bibitem{Lopez:2025arw}
F.~Lopez and N.~Bartolo, ``{Quantum signatures and decoherence during inflation
  from deep subhorizon perturbations},''
  \href{https://arxiv.org/abs/2503.23150}{{\ttfamily arXiv:2503.23150
  [astro-ph.CO]}}.

\bibitem{Pattison:2017mbe}
C.~Pattison, V.~Vennin, H.~Assadullahi, and D.~Wands, ``{Quantum diffusion
  during inflation and primordial black holes},''
  \href{https://dx.doi.org/10.1088/1475-7516/2017/10/046}{{\em JCAP} {\bfseries
  10} (2017) 046}, \href{https://arxiv.org/abs/1707.00537}{{\ttfamily
  arXiv:1707.00537 [hep-th]}}.

\bibitem{Ezquiaga:2018gbw}
J.~M. Ezquiaga and J.~Garc\'\i{}a-Bellido, ``{Quantum diffusion beyond
  slow-roll: implications for primordial black-hole production},''
  \href{https://dx.doi.org/10.1088/1475-7516/2018/08/018}{{\em JCAP} {\bfseries
  08} (2018) 018}, \href{https://arxiv.org/abs/1805.06731}{{\ttfamily
  arXiv:1805.06731 [astro-ph.CO]}}.

\bibitem{Biagetti:2018pjj}
M.~Biagetti, G.~Franciolini, A.~Kehagias, and A.~Riotto, ``{Primordial Black
  Holes from Inflation and Quantum Diffusion},''
  \href{https://dx.doi.org/10.1088/1475-7516/2018/07/032}{{\em JCAP} {\bfseries
  07} (2018) 032}, \href{https://arxiv.org/abs/1804.07124}{{\ttfamily
  arXiv:1804.07124 [astro-ph.CO]}}.

\bibitem{Figueroa:2020jkf}
D.~G. Figueroa, S.~Raatikainen, S.~Rasanen, and E.~Tomberg, ``{Non-Gaussian
  Tail of the Curvature Perturbation in Stochastic Ultraslow-Roll Inflation:
  Implications for Primordial Black Hole Production},''
  \href{https://dx.doi.org/10.1103/PhysRevLett.127.101302}{{\em Phys. Rev.
  Lett.} {\bfseries 127} no.~10, (2021) 101302},
  \href{https://arxiv.org/abs/2012.06551}{{\ttfamily arXiv:2012.06551
  [astro-ph.CO]}}.

\bibitem{Achucarro:2021pdh}
A.~Achucarro, S.~Cespedes, A.-C. Davis, and G.~A. Palma, ``{The hand-made tail:
  non-perturbative tails from multifield inflation},''
  \href{https://dx.doi.org/10.1007/JHEP05(2022)052}{{\em JHEP} {\bfseries 05}
  (2022) 052}, \href{https://arxiv.org/abs/2112.14712}{{\ttfamily
  arXiv:2112.14712 [hep-th]}}.

\bibitem{Cai:2022erk}
Y.-F. Cai, X.-H. Ma, M.~Sasaki, D.-G. Wang, and Z.~Zhou, ``{Highly non-Gaussian
  tails and primordial black holes from single-field inflation},''
  \href{https://dx.doi.org/10.1088/1475-7516/2022/12/034}{{\em JCAP} {\bfseries
  12} (2022) 034}, \href{https://arxiv.org/abs/2207.11910}{{\ttfamily
  arXiv:2207.11910 [astro-ph.CO]}}.

\bibitem{Animali:2022otk}
C.~Animali and V.~Vennin, ``{Primordial black holes from stochastic
  tunnelling},'' \href{https://dx.doi.org/10.1088/1475-7516/2023/02/043}{{\em
  JCAP} {\bfseries 02} (2023) 043},
  \href{https://arxiv.org/abs/2210.03812}{{\ttfamily arXiv:2210.03812
  [astro-ph.CO]}}.

\bibitem{Hooshangi:2023kss}
S.~Hooshangi, M.~H. Namjoo, and M.~Noorbala, ``{Tail diversity from
  inflation},'' \href{https://dx.doi.org/10.1088/1475-7516/2023/09/023}{{\em
  JCAP} {\bfseries 09} (2023) 023},
  \href{https://arxiv.org/abs/2305.19257}{{\ttfamily arXiv:2305.19257
  [astro-ph.CO]}}.

\bibitem{Vennin:2024yzl}
V.~Vennin and D.~Wands, {\em {Quantum Diffusion and~Large Primordial
  Perturbations from~Inflation}}.
\newblock 2025.
\newblock \href{https://arxiv.org/abs/2402.12672}{{\ttfamily arXiv:2402.12672
  [astro-ph.CO]}}.

\bibitem{Ozsoy:2023ryl}
O.~\"Ozsoy and G.~Tasinato, ``{Inflation and Primordial Black Holes},''
  \href{https://dx.doi.org/10.3390/universe9050203}{{\em Universe} {\bfseries
  9} no.~5, (2023) 203}, \href{https://arxiv.org/abs/2301.03600}{{\ttfamily
  arXiv:2301.03600 [astro-ph.CO]}}.

\bibitem{Tasinato:2020vdk}
G.~Tasinato, ``{An analytic approach to non-slow-roll inflation},''
  \href{https://dx.doi.org/10.1103/PhysRevD.103.023535}{{\em Phys. Rev. D}
  {\bfseries 103} no.~2, (2021) 023535},
  \href{https://arxiv.org/abs/2012.02518}{{\ttfamily arXiv:2012.02518
  [hep-th]}}.

\bibitem{Tasinato:2023ukp}
G.~Tasinato, ``{Large |\ensuremath{\eta}| approach to single field
  inflation},'' \href{https://dx.doi.org/10.1103/PhysRevD.108.043526}{{\em
  Phys. Rev. D} {\bfseries 108} no.~4, (2023) 043526},
  \href{https://arxiv.org/abs/2305.11568}{{\ttfamily arXiv:2305.11568
  [hep-th]}}.

\bibitem{Hillery:1983ms}
M.~Hillery, R.~F. O'Connell, M.~O. Scully, and E.~P. Wigner, ``{Distribution
  functions in physics: Fundamentals},''
  \href{https://dx.doi.org/10.1016/0370-1573(84)90160-1}{{\em Phys. Rept.}
  {\bfseries 106} (1984) 121--167}.

\bibitem{Habib:1992ci}
S.~Habib, ``{Stochastic inflation: The Quantum phase space approach},''
  \href{https://dx.doi.org/10.1103/PhysRevD.46.2408}{{\em Phys. Rev. D}
  {\bfseries 46} (1992) 2408--2427},
  \href{https://arxiv.org/abs/gr-qc/9208006}{{\ttfamily arXiv:gr-qc/9208006}}.

\bibitem{Serafini:2003ke}
A.~Serafini, F.~Illuminati, and S.~De~Siena, ``{Von Neumann entropy, mutual
  information and total correlations of Gaussian states},''
  \href{https://dx.doi.org/10.1088/0953-4075/37/2/L02}{{\em J. Phys. B}
  {\bfseries 37} (2004) L21},
  \href{https://arxiv.org/abs/quant-ph/0307073}{{\ttfamily
  arXiv:quant-ph/0307073}}.

\bibitem{landau1959fm}
L.~D. Landau, E.~M. Lifshitz, J.~B. Sykes, and W.~H. Reid, {\em Fluid
  Mechanics}.
\newblock Pergamon Press Oxford, England, 1959.

\bibitem{breuerTheoryOpenQuantum2002}
H.~P. Breuer and F.~Petruccione,
  \href{https://dx.doi.org/10.1093/acprof:oso/9780199213900.001.0001}{{\em The
  theory of Open Quantum Systems}}.
\newblock Oxford University Press, 2002.

\bibitem{Calzetta:2008iqa}
E.~A. Calzetta and B.-L.~B. Hu,
  \href{https://dx.doi.org/10.1017/CBO9780511535123}{{\em Nonequilibrium
  {Quantum} {Field} {Theory}}}.
\newblock Cambridge {Monographs} on {Mathematical} {Physics}. Cambridge
  University Press, Sept., 2008.

\bibitem{Boyanovsky:2015xoa}
D.~Boyanovsky, ``{Effective Field Theory out of Equilibrium: Brownian quantum
  fields},'' \href{https://dx.doi.org/10.1088/1367-2630/17/6/063017}{{\em New
  J. Phys.} {\bfseries 17} no.~6, (2015) 063017},
  \href{https://arxiv.org/abs/1503.00156}{{\ttfamily arXiv:1503.00156
  [hep-ph]}}.

\bibitem{Boyanovsky:2015tba}
D.~Boyanovsky, ``{Effective field theory during inflation: Reduced density
  matrix and its quantum master equation},''
  \href{https://dx.doi.org/10.1103/PhysRevD.92.023527}{{\em Phys. Rev. D}
  {\bfseries 92} no.~2, (2015) 023527},
  \href{https://arxiv.org/abs/1506.07395}{{\ttfamily arXiv:1506.07395
  [astro-ph.CO]}}.

\bibitem{Boyanovsky:2015jen}
D.~Boyanovsky, ``{Effective field theory during inflation. II. Stochastic
  dynamics and power spectrum suppression},''
  \href{https://dx.doi.org/10.1103/PhysRevD.93.043501}{{\em Phys. Rev. D}
  {\bfseries 93} (2016) 043501},
  \href{https://arxiv.org/abs/1511.06649}{{\ttfamily arXiv:1511.06649
  [astro-ph.CO]}}.

\bibitem{Burgess:2020tbq}
C.~P. Burgess, \href{https://dx.doi.org/10.1017/9781139048040}{{\em
  Introduction to {Effective} {Field} {Theory}}}.
\newblock Cambridge University Press, Dec., 2020.

\bibitem{Colas:2023wxa}
T.~Colas, {\em {Open Effective Field Theories for primordial cosmology :
  dissipation, decoherence and late-time resummation of cosmological
  inhomogeneities}}.
\newblock PhD thesis, Institut d'astrophysique spatiale, France, AstroParticule
  et Cosmologie, France, APC, Paris, 2023.

\bibitem{Salcedo:2024smn}
S.~A. Salcedo, T.~Colas, and E.~Pajer, ``{The open effective field theory of
  inflation},'' \href{https://dx.doi.org/10.1007/JHEP10(2024)248}{{\em JHEP}
  {\bfseries 10} (2024) 248},
  \href{https://arxiv.org/abs/2404.15416}{{\ttfamily arXiv:2404.15416
  [hep-th]}}.

\bibitem{Salcedo:2024nex}
S.~A. Salcedo, T.~Colas, and E.~Pajer, ``{An Open Effective Field Theory for
  light in a medium},'' \href{https://dx.doi.org/10.1007/JHEP03(2025)138}{{\em
  JHEP} {\bfseries 03} (2025) 138},
  \href{https://arxiv.org/abs/2412.12299}{{\ttfamily arXiv:2412.12299
  [hep-th]}}.

\end{thebibliography}
\end{document}